\documentclass[useAMS,usenatbib]{mnras}
\usepackage[latin1]{inputenc}
\usepackage[english]{babel}
\usepackage{graphicx}
\usepackage{natbib}
\usepackage{amsmath}
\usepackage{amsfonts}
\usepackage{amssymb}
\usepackage{amsbsy}
\title[Screening Mechanism and Late-time Cosmology : Role of a Chameleon-Brans-Dicke Scalar Field]{Screening Mechanism and Late-time Cosmology : Role of a Chameleon-Brans-Dicke Scalar Field}
\author[Soumya Chakrabarti, Koushik Dutta, Jackson Levi Said]{
Soumya Chakrabarti$^{1}$\thanks{E-mail : soumya1989@bose.res.in}, Koushik Dutta$^{2}$\thanks{E-mail : koushik@iiserkol.ac.in} and Jackson Levi Said$^{3}$\thanks{E-mail : jackson.said@um.edu.mt}
\\
$^{1}$Department of Theoretical Sciences, S. N. Bose National Centre for Basic Sciences, Kolkata, West Bengal 700106, India \\
$^{2}$Department of Physical Science, Indian Institute of Science Education and Research Kolkata, Mohanpur 741246, West Bengal, India \\
$^{3}$ Institute of Space Sciences and Astronomy, University of Malta, Msida, MSD 2080, Malta
}
\date{Accepted XXX. Received YYY; in original form ZZZ}
\pubyear{2022}
\begin{document}
\maketitle

\begin{abstract}
We discuss a way in which the geometric scalar field in a Brans-Dicke theory can evade local astronomical tests and act as a driver of the late-time cosmic acceleration. This requires a self-interaction of the Brans-Dicke scalar as well as an interaction with ordinary matter. The scalar field in this construct acquires a density-dependent effective mass much like a Chameleon field. We discuss the viability of this setup in the context of Equivalence Principle, Fifth Force and Solar System tests. The cosmological consistency is adjudged in comparison with observational data from recalibrated light-curves of type $Ia$ supernova (JLA), the Hubble parameter measurements (OHD) and the Baryon Acoustic Oscillation (BAO). We deduct that the astrophysical constraints indeed favour the existence of a mild scalar-matter interaction in the Jordan Frame.  
\end{abstract}


\section{Introduction}
The accelerated expansion of the universe is one of the most beautiful phenomena of modern era cosmology. The present phase of cosmic acceleration as well as the preceding epoch of deceleration receives quite conclusive support from a range of astrophysical observations \citep{perl1, riess1, melchi, sahni, jaffe, lange, halverson, netterfield, tonry, copeland}. As a matter of intrigue, no clear theoretical explanation can be found in the standard cosmological dynamics of a fluid description in the context of General Theory of Relativity (GR), inspite of the theory being the best theory of gravity till date. This requirement has driven physicists towards the foundation of a so-called Dark Energy (DE) component of the universe \citep{riess2} exhibiting a negative pressure that can drive the recent phase of cosmic acceleration. Astrophysical constraints also demand that the transition between subsequent epochs of the expanding universe must be smooth \citep{paddy1, paddy2}. Over the last few decades, a plethora of cosmological models are proposed in this regard. The simplest toy models are written using an energy component of the order of a Cosmological Constant or a time varying scalar field called the Quintessence \citep{zlatev, sahni}. The case of a Cosmological Constant has already become superfluous due to the ridiculous mismatch with quantum field theoretical estimates. On the other hand, the typical couplings between the quintessence field and the baryons seem to rule out a straightforward quintessence cosmology in view of the stringent constraints coming from the fifth-force experiments \citep{adelberger}. Nevertheless, using a quintessence field in a broader context can produce solutions of considerable interest. For instance, two interesting possibilities exist where the scalar field is either pseudo-Nambu-Goldstone-Boson (pNGB) \citep{frieman} or the scalar field hides its couplings via screening mechanisms \citep{khoury, hinterbichler}. In the latter case, it is assumed that the cosmological constant is zero due to some unknown reasons. For the case of a pNGB, the potential is natural from the view point of quantum field theory even though it faces some technical difficulties in UV complete theories \citep{dutta1, dutta2, banks, arkani}. \medskip

The present work is motivated primarily by scalar extended theories of gravity where the acceleration of the universe is driven by a geometric DE. Brans-Dicke (BD) theory arguably is the paradigm of any such extension \citep{bransdicke, misner} where the effective gravitational coupling can vary \citep{faraoni}. The BD scalar field $\phi$ alongwith a dimensionless parameter $\omega$ mark the characteristics of the theory in comparison with standard GR. Local astronomical tests in general demand strict constraints over this BD parameter $\omega$ and more often than not, this requirement poses a problem in finding consistent cosmological solutions \citep{nb1, faraoni1}. As a consequence, a standard BD theory, inspite of having immense initial potential can not quite be regarded as \textit{`the better theory'} of gravity. It can still provide excellent toy models though, in particular of the early acceleration of the universe in the form of an extended inflation \citep{guth, mathia, la} and a late time accelerated expansion \cite{nb2} as well. Addition of a self-interaction potential of the BD scalar field or an additional quintessence scalar field in the action provides some of the simplest extensions of the theory and these are studied in literature with rigorous details \cite{faraoni2, bertolami, sensen, nb3}. Making $\omega$ a function of the scalar field is another interesting possibility which can provide a revival of the standard BD theory \citep{bergmann, wagoner, nordtvedt, barker}. For more aspects of standard BD theory and possible extensions, we refer to some of the existing literature \citep{soti}. \medskip

We intend to focus on the role of scalar interactions in these extended theories. We find motivation in a theory where the scalar, apart from its self-interaction, has interaction with ordinary matter as well. We call the theory a generalized Chameleon-Brans-Dicke (CBD) theory and give it's basic formalism in both Einstein and Jordan frames, employing a suitable conformal transformation following the standard literature \citep{jarv, quiros}. In the process, the scalar interaction in the Einstein Frame is modified beyond the standard conformal coupling with matter and leads to modification of gravitational interaction in small scale as well as large scale structure formations. However, the solar system constraints and fifth force experiments require that such a scalar must remain subdued locally \citep{jain}, i.e., in high-density regions. In our setup, this is possible due to the dominant scalar-matter interaction in the solar system, which can easily decouple the scalar from matter. A scalar field acquiring such a nature is popular for the `Chameleon' like feature as in evasion from detection by local experiments \citep{will2}. The set of requirements for the chameleon nature of the scalar can generate a novel set of constraints over any scalar dominated theory of gravity \citep{khoury, gubser, upadhye, brax}. These constraints essentially ensure that the Equivalence Principle (EP) violation is avoided on the solar system scales \citep{will1, baessler, damour, huey, hill, ellis}. \medskip

Due to the generalized scalar interaction, the scalar behaves in an entirely different manner in the solar system even if it is quite massive around Earth. More importantly, on cosmological scales one can use such a scalar to construct an entity with a mass of the order of present value of Hubble parameter and provide a perfect candidate to generate the acceleration of the universe. Previous endeavors with similar motivation have produced promising cosmological solutions \citep{das, easson, mota1, mota2, dasbanerjee} as well as a possibility of interacting dark energy-dark matter models of cosmology \citep{zimdahl}, however, the equivalence principle requirements were never really considered. We show that the generalized CBD scalar model can satisfy these constraints alongwith the astrophysical requirements to generate a viable model of late time cosmology that can describe the smooth transition of the universe from deceleration into acceleration. We employ a simple statefinder diagnostic to discuss the cosmological issues. We estimate the constraints on model parameters with a statistical analysis with observational data sets such as the Joint Light-Curve Analysis (JLA) \citep{betoule, simon, stern, chuang, moresco, blake, delubac}, Hubble parameter estimation (OHD) \citep{ade} and the Baryon Accoustic Oscillation (BAO) \citep{beutler, anderson} data, using a Markov Chain Monte Carlo (MCMC) numerical simulation \citep{mcmc}. \medskip

However, demanding that the Milky Way galaxy is screened, it has recently been proved that standard chameleon like theories need to satisfy $m_o^{-1} \lesssim \text{Mpc}$, where $m_0$ is the mass of the chameleon field at present cosmological density \citep{wang}. With the assumption of a vanishing divergence of the matter stress-energy tensor, this essentially leads to the deduction that a standard chameleon field alone can not be the driver of the present cosmic acceleration. We emphasize that the BD scalar in this theory interacts with standard (baryonic) matter in the Jordan Frame itself, such that the standard matter conservation and the geodesic equations are altered. The steep constraint over the mass of the chameleon field at the cosmological density therefore, does not apply straightaway. \medskip

In the next section, Sec. II, we discuss the basic setup of the theory and the transformation in between Jordan and Einstein Frame. This is followed by a discussion of static solutions of spherical bodies such as the Earth in this theory, in Sec. III. Subsequently we analyze these solutions and comment on the fifth force and solar system constraints on the profile of the earth, in Sec. IV. Sec. V contains a discussion on late-time cosmology from a kinematic approach and thereafter we conclude in Sec. VI. We also include some detailed calculations in the Appendix. 

\section{Basic Setup : Action and The Equations}
The generalized Chameleon-Brans-Dicke (CBD) action in Jordan frame is written as
\begin{eqnarray}\label{1}\nonumber
&& S = \frac{1}{16\pi}\int d^4x \sqrt{-\bar{g}} \{\phi \bar{R}-\frac{\omega_{BD}}{\phi}\bar{g}^{\mu\nu}\bar{\nabla}_{\mu}\phi\bar{\nabla}_{\nu}\phi - V(\phi) \\&& 
+ 16\pi f(\phi)L_m\},
\end{eqnarray} 
where $\bar{R}$ is the Ricci scalar and $\phi$ is the BD scalar field in the Jordan frame. $\omega_{BD}$ is the dimensionless BD parameter. We have two unknown analytical functions of $\phi$ at the outset, $V(\phi)$ and $f(\phi)$ and they represent the self-interaction and the scalar-matter interaction respectively. To write the action in Einstein Frame we use a conformal transformation
\begin{eqnarray}\label{a2}
&&\bar{g}_{\mu\nu}\rightarrow g_{\mu\nu}=\Omega^2 \bar{g}_{\mu\nu}, \\&&
\Omega = \sqrt{G \phi}.
\end{eqnarray} 
We also redefine the BD scalar as
\begin{equation}\label{a3}
\varphi(\phi) = \sqrt{\frac{2\omega_{BD}+3}{16\pi G}} \ln (\frac{\phi}{\phi_0}).
\end{equation}
One needs to ensure that $\omega_{BD} > -\frac{3}{2}$. With these, the Einstein Frame action becomes
\begin{eqnarray}\label{a5}\nonumber
&& S_{EF} = \int d^{4}x \sqrt{-g} \{\frac{R}{16\pi G} -\frac{1}{2}g^{\mu\nu}\nabla_{\mu}\varphi\nabla_{\nu}\varphi -U(\varphi) \\&&
+ \exp(-\frac{\sigma\varphi}{M_p})~f(\varphi) L_{m}\}.
\end{eqnarray}
The parameter $\sigma$ carries the original BD parameter, defined as 
\begin{equation}
\sigma = 8\sqrt{\frac{\pi}{2\omega_{BD}+3}},
\end{equation}
and therefore, constraints on $\sigma$ essentially leads one to constraints over $\omega_{BD}$. The self-interaction of the BD field in Einstein frame is
\begin{equation}\label{a6}
U(\varphi) = V(\phi(\varphi))~\exp(-\sigma\varphi/M_{p}).
\end{equation}
Variation of the Einstein frame action in Eq. (\ref{a5}) with respect to the metric and the redefined scalar field leads one to the field equations
\begin{equation}\label{2}
G_{\mu\nu} = M_p^{-2} (h(\varphi)T^{m}_{\mu\nu} + T^{\varphi}_{\mu\nu}),
\end{equation}
and
\begin{equation}\label{3}
\Box \varphi - U'(\varphi) = -h'(\varphi) L_{m}. 
\end{equation}

Note that a redefined interaction as in $h(\varphi) = e^{-\sigma\varphi/M_{p}}f(\varphi)$ dictates the field equations, as in the scalar evolution equation there is now a new source term. The same can be realized with Jordan frame field equations as well. In the rest of the manuscript, we analyse static and time-evolving solutions of the field equations to make some important deductions. 

\section{The Brans-Dicke scalar as a Chameleon}
In this section we discuss that the geometric scalar in the present setup can exhibit a chameleon like behavior by virtue of it's interactions. The standard chameleon mechanism and the requirements to satisfy the fifth force constraints are discussed in details in the works of \citep{khoury, waterhouse}. Realisation of this mechanism requires the static profile of the scalar in Einstein frame. The self-interaction potential $U(\varphi)$ must be taken in a runaway form which also brings in additional motivations from string theory \citep{huey, hill, ellis, barreiro, binetruy, barrow} and standard quintessence cosmological models \citep{zlatev, sahni}. We recall that the equation of motion for $\varphi$ is

\begin{equation}\label{3cham}
\Box \varphi = U'(\varphi) - h'(\varphi) L_m.
\end{equation}

The scalar-matter energy exchange and the field equations are directly related to the choice of matter Lagrangian density $L_m$ \citep{schutz, brown}. In literature, two well-discussed choices remain, $L_m = p_m$ and $L_m = -\rho_m$ \citep{bertolami1, bertolami2} and it has been proved that if the scalar-matter interaction is negligible, these choices are equivalent \citep{bertolami3, bertolami4, sotifara}. It is not a straightforward case in presence of interaction and the question of equivalence already inspires qualitative arguements over the choice of matter lagrangian \citep{faraonimatter}. Avoiding the conundrum, we straightaway take $L_m = -\rho_m$ and write the RHS of Eq. (\ref{3cham}) as an effective potential
\begin{equation}\label{effpot2}
U_{eff}(\varphi) = U(\varphi) - \rho_{0} \int \left(\frac{\sigma}{M_p} - \frac{df}{d\varphi}\right)e^{\frac{\sigma \varphi}{2M_p}}.
\end{equation}
$\rho_m \equiv {\rho}_{0}e^{\frac{3\sigma \varphi}{2M_p}}$ is the energy density conserved in Einstein frame (see for instance \citep{waterhouse} on more discussions in this regard). $U(\varphi)$ is monotonic as in a runaway form but $U_{eff}$ can have a minimum. We assume that $\varphi_{min}$ is the value of $\varphi$ at the minimum such that

\begin{equation}
U_{{eff}{,\varphi}(\varphi_{min})} = 0\,.
\label{phimin}
\end{equation}

Similarly, we define the mass of small fluctuations about $\varphi_{min}$ as
\begin{equation}
m^{2}_{min} = U_{{eff}{,\varphi,\varphi}(\varphi_{min})} = 0\,.
\label{mmin}
\end{equation}

Both $\varphi_{min}$ and $m_{min}$ are functions of density and this has notable consequences. For a simple spherical mass distribution of radius $R$, the minima of $U_{eff}$ is expected, for instance $\varphi_c$ for $r < R$ and $\varphi_\infty$ for $r > R$. In a similar fashion, the mass of small fluctuations in the respective regions are $m_c$ and $m_\infty$. Thus the static scalar profile outside the sphere depends on the size of the sphere. While small objects can only produce mild perturbation on the $\varphi$ profile, for large objects, it relies on the solution of Eq. (\ref{3cham}), the nature of the effective potential in Eq. (\ref{effpot2}) and the smooth matching of $\varphi$ solutions inside and outside. We have discussed the solutions in some details in the later parts of the manuscript. For simplicity, it is fruitful to assume at the outset that
\begin{equation}
m_\infty R \ll 1,
\end{equation}
and think of the spherical object as an accumulation of infinitesimal disks or volume elements $dV$. This means that one has to solve the scalar Klein-Gordon equations for these thin disks and assemble to write the distribution of $\varphi$. Intriguingly, it can be proved that only the volume elements lying within a thin shell of thickness $\Delta R$ close to the surface of the sphere dominates this contribution (The scalar profiles for other disks inside the sphere contributes proportional to $\exp(-m_c\tilde{r})$ and are subdued, since $\varphi \approx \varphi_c$ and $m_c \gg m_\infty$. The solution for this thin shell volume elements of thickness $\Delta R$ near the surface can be written as
\begin{equation}\label{thinsoln}
\varphi(r) \approx -\left(\frac{\beta_{0}}{4\pi M_{P}}\right)\left(\frac{3\Delta R}{R}\right)\frac{M_c e^{-m_\infty r}}{r} + \varphi_\infty.
\end{equation}

The radial distribution of a chameleon field essentially is driven by the thin shell satisfying

\begin{equation}
\frac{\Delta R}{R}\ll 1 .
\label{thincond}
\end{equation}

However, this is only true for large spherical objects such as the Earth. For small objects, picking out a thin shell close to the boundary surface is not possible as $\Delta R/R > 1$ and therefore the entire volume contributes to the $\varphi$-field outside. For them the whole thing acts like a thick shell and the exterior solution can be found as

\begin{equation}\label{warmup}
\varphi(r) \approx -\left(\frac{\beta_{0}}{4\pi M_{P}}\right)\frac{M_c e^{-m_\infty r}}{r} + \varphi_\infty.
\end{equation}

These two solutions in Eqs. (\ref{thinsoln}) and (\ref{warmup}) differ by the thin-shell suppression factor of $\Delta R_c/R_c$. To mathematically derive the chameleon profile, we consider that the sphere of radius $R$ has homogeneous density $\rho_c$ and is submerged in a medium of homogeneous density $\rho_\infty$, very much like any astronomical body in our solar system. Then Eq. (\ref{3cham}) becomes

\begin{eqnarray}
&&\frac{d^2\varphi}{dr^2} + \frac{2}{r}\frac{d\varphi}{dr} = U_{,\varphi} + \rho_{0}(r) \frac{\beta(\varphi)}{M_p}e^{\frac{\sigma \varphi}{2M_p}}, \\&&
\frac{\beta(\varphi)}{M_p} = \left(\frac{df}{d\varphi} - \frac{\sigma}{M_p}\right),
\label{sun2}
\end{eqnarray}
where
\begin{equation}
\rho_{0} (r) = \left\{
\begin{matrix}
\rho_c\qquad {\rm for}\;\;\; r < R \cr
\rho_\infty\qquad {\rm for}\;\;\; r > R
\end{matrix}
\right. \,.
\label{rho}
\end{equation}

The trick is to approximate the effective potential according to physical requirements in separate regions and write a solution for the scalar in closed form. Outside the sphere the effective potential pushes the scalar towards $\varphi_\infty$ and we use a damped harmonic oscillator approximation \citep{waterhouse} to write, for $r > R$,

\begin{equation}
\frac{d^2\varphi}{dr^2}+\frac{2}{r}\frac{d\varphi}{dr} = m_{\infty}^{2}\left(\varphi - \varphi_{\infty}\right).
\end{equation}

One can write the general solution to this as
\begin{equation}
\varphi(r) = A\frac{e^{-m_{\infty}(r-R)}}{r} + B\frac{e^{m_{\infty}(r-R)}}{r} + \varphi_{\infty}.
\label{harmonicsolution}
\end{equation}

The dimensionless parameters $A$ and $B$ can be determined using the asymptotic condition $\varphi \rightarrow \varphi_\infty$ as $r \rightarrow \infty$. This leads to $B = 0$ and one can write
\begin{equation}
\varphi(r) = A\frac{e^{-m_\infty(r-R)}}{r} + \varphi_\infty.
\label{outsidesolution}
\end{equation}

As discussed earlier, the interior solution is the assembly of two different approximations in two different regions, the thin shell close to the surface and the rest, separated at some $r = R_{c}$. For $\left[0 , R_{c}\right]$ the chameleon $\varphi \sim \varphi_{c}$ and for $\left[R_{c} , R\right]$, $\varphi \gg \varphi_{c}$. 

\begin{enumerate}
\item{Approximation $1$:  $\varphi \gg \varphi_{c}$. \\
This approximationi is valid in the region where the chameleon field has moved away from it's minima. Here, $\sim e^{\sigma\varphi/M_\text{p}}$ dominates and the runaway potential $U(\varphi)$ decays rapidly. Assuming $\varphi \ll M_\text{p}$ the effective potential then becomes 
\begin{equation}
U_{{eff},\varphi} \left (\varphi\right) \approx \frac{\beta(\varphi)}{M_\text{p}}\rho_{c}.
\label{linearapproximation}
\end{equation}
Eq. (\ref{sun2}) then takes the form
\begin{equation}\label{approximate_eq}
\frac{d^2\varphi}{dr^2} + \frac{2}{r}\frac{d\varphi}{dr} \approx \frac{\beta(\varphi)}{M_\text{p}}\rho_{c}.
\end{equation}

We note that, unlike the case of a standard chameleon \citep{khoury}, the present case sees a modification due to the function $\beta(\varphi)$. This is directly related to the scalar-matter interaction in Jordan frame $f(\varphi)$. As a simple example and to remain as close to the original chameleon setup as possible, we choose $f(\varphi)$ to be a slowly varying function of $\varphi$ 
\begin{equation}\label{fphi1}
f(\varphi) = f_{1} + f_{0}\varphi.
\end{equation}
Therefore
\begin{equation}\label{fphi2}
\beta(\varphi) = \left(f_{0}M_{p} - \sigma \right) = \beta_{0}.
\end{equation}

With this, the solution to Eq. (\ref{approximate_eq}) can be written as
\begin{equation}
\varphi\left(r\right) = \frac{\beta_{0}}{6M_{p}}\rho_{c}r^{2} + \frac{C}{r} + D\varphi_{c}.
\label{inharmonicsolution}
\end{equation}
}
\item{Approximation 2 : $\varphi \sim \varphi_\text{c}$. \\
In this limit we use a damped harmonic oscillator approximation to write
\begin{equation}
V_{\text{eff}}\left(\varphi\right)\approx m_\text{c}^2\left(\varphi - \varphi_\text{c}\right),
\label{harmonicapproximation}
\end{equation}
whose solution is
\begin{equation}
\varphi(r) = E\frac{e^{-m_\text{c}r}}{r} + F\frac{e^{m_\text{c}\left(r-R_\text{c}\right)}}{r} + \varphi_\text{c}.
\label{otherharmonicsolution}
\end{equation}
}
\end{enumerate}

A complete solution to Eq. (\ref{sun2}) requires a smooth matching of these two solutions. The process involves lengthy calculations and we include most details in the Appendix. Of all the possible cases, we particularly concentrate on the $R_\text{c} = 0$ case, or the thick-shell solution and the $0 < R_\text{c} < R$, or the thin-shell solution. Together, these two cases enable us to comment on the thickness of a spherical object in the gravitational field. The exterior approximate solution for all the cases is 
\begin{displaymath}
\varphi\left(r\right) = A\frac{e^{-m_\infty(r-R)}}{r} + \varphi_\infty.
\end{displaymath}
The dimensionless constant $A$ is of utmost importance as it gives us the strength of the chameleon force due to the scalar (see Appendix for more details). $A$ is determined in term of the known physical parameters of the lagrangian. The final solutions for the thick shell case and the thin shell case with explicit expressions of $A$ are found as

\begin{eqnarray}
&&\varphi_\text{thick}\left(r\right) \sim - \frac{\beta_{0}}{4\pi M_\text{p}}\left(\frac{4}{3}\pi R^3 \rho_\text{c}\right)\frac{e^{-m_\infty(r-R)}}{r} + \varphi_\infty \\&&\nonumber
\varphi_\text{thin}\left(r\right) \sim - \frac{\beta_{0}}{4\pi M_\text{p}}\left(\frac{4}{3}\pi R^3\rho_\text{c}\right)\left(3\frac{M_\text{p}\left(\varphi_\infty - \varphi_\text{c}\right)}{\beta_{0}\rho_\text{c}R^2}\right) \\&&
\frac{e^{-m_\infty(r-R)}}{r} + \varphi_\infty.
\end{eqnarray}

The factor $3\Delta R/R$ is found through the relation
\begin{equation}\label{thinshellc}
\frac{\Delta R}{R}\equiv\frac{M_\text{p}\left(\varphi_\infty - \varphi_\text{c}\right)}{\beta_{0}\rho_\text{c}R^2}.
\end{equation}
Using this we define a chameleon suppression factor (see for instance Eq. (\ref{thinshellfactor}) in Appendix)
\begin{align*}
W&\equiv -A\left[\frac{\beta_{0}}{4\pi M_\text{p}}\left(\frac{4}{3}\pi R^3\rho_\text{c}\right)\right]^{-1}\\
&=-A\frac{3M_\text{p}}{\beta_{0} R^3\rho_\text{c}}.
\end{align*}

The suppression factor is directly responsible for bringing the term $\beta_{0}$ into play, resulting in a modification of the standard chameleon screening. The generalized scalar-interaction in this process makes sure that all the results include the additional function $\frac{\beta(\varphi)}{M_p} = \left(\frac{\sigma}{M_p} - \frac{df}{d\varphi}\right)$. A slowly varying approximation $\beta(\varphi) = \left(f_{0}M_{p} - \sigma \right) = \beta_{0}$ essentially leads to a straightforward scaling of the original results and keeps the theory close to the original chameleon theory. The suppression factor lets us comment on the thickness of an astronomical object (e.g. the earth) and plays a crucial part in determining the evasion of the scalar from astronomical tests.

\section{Constraints from Fifth Force, EP Violation and Solar System Tests}
The primary requirement is that the chameleon field must not be detected in any laboratory vacuum experiments. To have an accord with this we take the mass of the chameleon field inside a chamber to be of the order of $R_\text{vac}^{-1}$ (for more discussions see \citep{khoury}), where the radius of the vacuum chamber is $R_\text{vac}$.
\begin{equation}
m_{vac} \equiv \sqrt{U_{,\varphi\varphi}(\varphi_{vac})} = R_{vac}^{-1}\,.
\label{mv}
\end{equation}

For a more complicated system as in an Earth $+$ atmosphere system, we combine the solutions mentioned in the previous section for different regions, labelled by their density and solve for an increased number of continuity conditions. For the sake of brevity we only focus on the results as the calculations are just an extended case of the solutions provided in Appendix.

\subsection{Profile for the Earth} \label{earth}
To write the radial distribution of the scalar in and around earth, we write the Earth-Atmosphere system as a spherical distribution of three phases of matter density
\begin{equation}
\rho (r) = \left\{
\begin{matrix}
\;\;\;\rho_\oplus\qquad\;\;\; {\rm for}\;\;\; 0<r<R_\oplus \cr
\;\;\;\;\;\;\;\;\;\;\rho_{atm}\qquad {\rm for}\;\;\; R_\oplus< r <R_{atm} \cr
\rho_G\qquad\;\;\; {\rm for}\;\;\; r>R_{atm}
\end{matrix}
\right.
\label{rhoearth}
\end{equation}
$\varphi_\oplus$, $\varphi_{atm}$ and $\varphi_G$ are the field values minimizing the effective potential in the three layers of the density distribution. Similarly, $m_\oplus$, $m_{atm}$ and $m_G$ are the respective masses of fluctuation. It is necessary for the atmosphere to have a thin shell in order to avoid large violations of the EP as extensively proved by \citep{khoury}. In this scenario, the scalar value is supposed to remain around the stable minimum $\varphi_{atm}$ in the atmosphere. Moreover, assuming that the Earth has a thin shell, $\varphi \approx \varphi_\oplus$ inside the Earth. Then from Eq. (\ref{thinshellc}), the thin-shell condition for the atmosphere can be written as
\begin{equation}
\frac{\Delta R_{atm}}{R_{atm}} = \frac{\varphi_G - \varphi_{atm}}{6\beta_{0} M_{P}\Phi_{atm}} \ll 1\,,
\end{equation}
where $\Phi_{atm} \equiv \rho_{atm}R_{atm}^2/6M_{P}^2$. If the atmosphere must have a thin shell the shell must atleast be thinner than the the atmosphere itself. Therefore, taking $R_{atm}\equiv R_\oplus +$ 10 km, $\Delta R_{atm}/R_{atm} \;\lesssim\; 10^{-3}$. Taking $\rho_{atm}\approx 10^{-4}\rho_{\oplus}$ and $\Phi_{atm}\approx 10^{-4}\Phi_\oplus$, we write the condition for the atmosphere to have a thin shell as

\begin{equation}
\frac{\Delta R_\oplus}{R_\oplus} \equiv \frac{\varphi_G - \varphi_{atm}}{6\beta_{0} M_{P}\Phi_{\oplus}} < 10^{-7} \,.
\label{condatm}
\end{equation}

\subsection{Fifth Force Searches} 
The potential energy associated with the fifth force between test bodies $M_1$ and $M_2$ in a separation of $r$ is defined as

\begin{equation}
V(r) =  - \alpha \frac{M_1M_2}{8\pi M_{P}^2}\frac{e^{-r/\lambda}}{r}.
\label{fifthpot}
\end{equation}

$\alpha$ and $\lambda$ are the strength and range of the interaction. We have already taken the range of chameleon originated interactions in vacuum to be of the order of the size of a vacuum chamber, i.e., $\lambda \approx R_{vac}$. Assumming the scale of $\lambda$ to be centimetres, the widely accepted bound on $\alpha$ is given by standard laboratory experiments 
\begin{equation}
\alpha < 10^{-3}\,. 
\label{hos}
\end{equation}

In the present case, we proceed by considering two identical test bodies of uniform density $\rho_c$, radius $R_c$ and total mass $M_c$ and claim that the test masses must have a thin shell satisfying the criterion
\begin{equation}
\frac{\Delta R_c}{R_c}\equiv \frac{\varphi_{vac} - \varphi_c}{6\beta_{0} M_{P}\Phi_c} \ll 1 \,.
\label{condbodyvac}
\end{equation}
If the test masses have a thin shell, the chameleon field profile is written as
\begin{equation}
\varphi(r) \approx -\left(\frac{\beta_0}{4\pi M_{P}}\right)\left(\frac{3\Delta R_c}{R_c}\right)\frac{M_c e^{-r/R_{vac}}}{r} + \varphi_{vac}\,,
\label{bodyvac}
\end{equation}
with the corresponding potential energy as
\begin{equation}
V(r) = -2\beta_{0}^{2}\left(\frac{3\Delta R_c}{R_c}\right)^2
\frac{M_c^2}{8\pi M_{P}^2}\frac{e^{-r/R_{vac}}}{r} \,.
\end{equation}
Comparing with Eq. (\ref{fifthpot}), one can write the bound in Eq.~(\ref{hos}) as
\begin{equation}
2\beta_{0}^{2}\left(\frac{3\Delta R_c}{R_c}\right)^2 \;\lesssim\; 10^{-3}\,,
\label{condbodyvac2}
\end{equation}

which essentially gives an idea on how the $\beta_{0}$ parameter must be chosen and this leads to an immediate idea on how strong the scalar-matter interaction can be in the generalized CBD construct.

\subsection{Solar System Tests} \label{solar}
An important constraint comes by comparison of free-fall acceleration in the Moon-Earth-Sun system, measured by laser ranging experiments \citep{will1}.
\begin{equation}
\frac{|a_{Moon} - a_\oplus|}{a_N} \;\lesssim\; 10^{-13}.
\label{diffacc}
\end{equation}
$a_N$ is the Newtonian acceleration. The Sun, Earth and Moon and their respective atmospheres are spherical and all subject to the thin shell effect. It is obvious for the Sun to exhibit thin-shell since it has a Newtonian potential much larger than the Earth. For the Moon, assuming $\varphi_G \gg \varphi_{Moon}$ and using Eq. (\ref{condatm}) this condition can be written as
\begin{equation}
\frac{\Delta R_{Moon}}{R_{Moon}} \sim \frac{\Delta R_\oplus}{R_\oplus}\frac{\Phi_\oplus}{\Phi_{Moon}} < 10^{-5}\,,
\end{equation}
where standard values of $\Phi_\oplus = 10^{-9}$ and $\Phi_{Moon} = 10^{-11}$ are used. Now, studying the $\varphi$ profiles outside each spherical body one can write explicitly the acceleration of the Earth and Moon towards the Sun. Comparison of them produces a constraint over the free-fall acceleration as 
\begin{equation}\label{diffacc1}
\frac{|a_{Moon} - a_\oplus|}{a_N} \approx \beta_{0}^2 \left(\frac{\Delta R_\oplus}{R_\oplus}\right)^2 < \beta_{0}^2\cdot 10^{-14}\,.
\end{equation}
This is a scaled version of the constraint in Eq. (\ref{diffacc}), courtesy of the scalar-matter interaction parameter $\beta_{0}$.

\subsection{Constraints on the Scalar Interactions and the Brans-Dicke Parameter} \label{condmass}
The constraints on different model parameters induced by the overall requirements of fifth force, EP violation and solar system constraints are discussed in this section. The model parameters come from the scalar self-interaction, the scalar-matter interaction and the Brans-Dicke parameter. The BD scalar self-interaction is written as a power-law potential of runaway form
\begin{equation}
U(\varphi) = M^{4+n}\varphi^{-n}.
\label{Vtypical}
\end{equation}
$n > 0$ and $M$ is of mass dimension. The essential constraint we focus upon is Eq. (\ref{condatm}), which gives
\begin{equation}
\frac{\Delta R_\oplus}{R_\oplus} \equiv \frac{\varphi_G-\varphi_{atm}}{6\beta_{0} M_{P}\Phi_{\oplus}} < 10^{-7}.
\label{condition2}
\end{equation}

For $\varphi = \varphi_G$, the minimized effective potential and $\rho = \rho_G$ we can write
\begin{equation}
U_{,\varphi}(\varphi_G) + \beta_{0} \rho_G e^{\sigma\varphi_{G}/M_{P}}/{2M_{P}} = 0.
\end{equation}

With the runaway potential and an assumption of $\sigma\varphi_{G}/M_{P} \ll 1$ this leads to

\begin{equation}
\varphi_G = \left(\frac{n M^{4+n} M_{P}}{\beta_{0}\rho_G}\right)^{\frac{1}{(n+1)}}\,.
\label{phiGsolved}
\end{equation}

If one takes $\rho_G = 10^{-24}\;{\rm g}/{\rm cm}^3$ and $\Phi_\oplus=10^{-9}$, Eq.~(\ref{condition2}) leads to a bound on $M$ as
\begin{equation}
M\;<\;\left(\frac{6^{n+1}}{n}\right)^{\frac{1}{n+4}}\beta_{0}^{\frac{n+2}{n+4}}\cdot 10^{\frac{15n-7}{n+4}}\cdot (1\;{\rm mm})^{-1}\,.
\label{condM1}
\end{equation}

The condition in Eq. (\ref{condM1}) is the main constraint to be derived before we move on to cosmology. As an example, if $n, \beta_0 \sim O(1)$ Eq. (\ref{condM1}) leads one to a constraint on $M$ such that $M \lesssim 10^{-3}$ eV. Remarkably, this is comparable with the mass scale of cosmological constant. It is a hint, if only tiny, that under suitable conditions, a chameleon scalar can drive the late-time acceleration to a fair degree of consistency. \medskip

Another important fact is, the choice of $\beta_{0} \sim {\it O}(1)$ leads to a constraint on the couplng parameter $\sigma$. We also recall that $\sigma$ carries the original BD parameter, defined as $\sigma = 8\sqrt{\frac{\pi}{2\omega_{BD}+3}}$. Together with Eqs. (\ref{fphi1}) and (\ref{fphi2}) this constraint becomes

\begin{equation}\label{resultt1}
f_{0}M_{p} - 8\left( \frac{\pi}{2\omega_{BD}+3}\right)^{1/2} \sim {\it O}(1),
\end{equation}
or alternatively
\begin{equation}\label{omegabdexp}
\omega_{BD} \sim \frac{32\pi}{\left(f_{0}M_{p} - 1\right)^{2}} - \frac{3}{2}.
\end{equation}

Clearly the parameter $f_{0}$ must be of the dimension of $M_{p}^{-1}$. We also stress upon the fact that a slowly varying $f(\varphi)$ as in Eq. (\ref{fphi1}) is perhaps the most suitable choice of the scalar-matter chameleon interaction, since one is working under the precinct of $\varphi \ll M_\text{p}$. For instance, if $f(\varphi) \propto f_{1} + f_{0} \varphi^{n}$ with $n \geq 3$, $\beta(\varphi)$ is then of the order of $\varphi^{2}$ or an even more sharply increasing function of $\varphi$ (courtesy of Eq. (\ref{sun2})). Under the assumption of $\varphi \ll M_\text{p}$, this ultimately nullifies $\rho_{0}(r) \frac{\beta(\varphi)}{M_p}e^{\frac{\sigma \varphi}{2M_p}}$ in the effective potential, leaving behind $U_{,\varphi}$. In case of an exponential $f(\varphi)$, one can follow similar arguements, as for $f(\varphi) \sim f_{0}e^{\frac{f_{1}\varphi}{M_p}}$, the exponential factor is suppressed and $\beta(\varphi) \sim \beta_{1} = (f_{0}f_{1} - \sigma)$. In such a case, one can easily calculate that the constraint on the Brans-Dicke parameter becomes
\begin{equation}
\omega_{BD} \sim \frac{32\pi}{\left(f_{0}f_{1} - 1\right)^{2}} - \frac{3}{2}.
\end{equation}
These constraints primarily help us determine the scale of the model parameters such as $f_{0}$, $f_{1}$, which signify the strength of interactions. Aditionally it also hints us that perhaps with a suitable scalar-matter interaction, it is also possible to realize a viable theoryo of gravity wth relaxed constraints on the BD parameter.

\section{Late time cosmology with a Chameleon-Brans-Dicke scalar field}
It is an interesting proposition to look into the cosmological nature of a Brans-Dicke scalar field of chameleon nature \citep{Noller:2018wyv, 2011JCAP...08..004S, Yang:2018xah, Bernardo:2021qhu}. In the present case, the scalar field already appears to pass smaller scale phenomenology tests. We take a flat homogeneous and isotropic cosmology for our Brans-Dicke Chameleon scalar field. We start by writing the Friedmann equations in the Jordan frame for this background cosmology, giving
\begin{align}\label{cosmofield}
    3H^2 &= \frac{\rho_m f}{\phi} + \frac{\omega}{2}\frac{\dot{\phi}^2}{\phi^2} - 3H\frac{\dot{\phi}}{\phi}\,,\\
    2\dot{H} + 3H^2 &= -\frac{\omega}{2}\frac{\dot{\phi}^2}{\phi^2} - \frac{\ddot{\phi}}{\phi} - 2H\frac{\dot{\phi}}{\phi}\,,
\end{align}
and the Klein-Gordon equation
\begin{equation}
    \left(2\omega+3\right)\left(\ddot{\phi} + 3H\dot{\phi}\right) = \rho_m f + \rho_m f' \phi\,,
\end{equation}
where $H(t)$ is the Hubble parameter and where dots denote derivatives with cosmic time and primes derivatives with the scalar field. Also, we recall that $f$ is described through Eq.~(\ref{fphi1}) and that $\omega$ is expressed through Eq.~(\ref{omegabdexp}). On the matter side, the energy density $\rho_m$ may have an impact on the effects of dark matter but a fuller galactic rotation curve analysis would need to be carried out to fully understand how this would play out. As usual, we take the matter contribution to be pressure-less dust. Using the the independent equations above, we can write the matter conservation equation giving the integrated form
\begin{equation}
    \rho_m = \frac{\rho_0}{a^3 f^{3/2}}\,,
\end{equation}
where $\rho_0$ is an integration constant. This expression again reiterates that for an interacting CBD theory, the matter conservation equation is modified by the scalar-matter interaction terms. 
\begin{itemize}
\item {First, we incorporate a direct implementation of Markov chain Monte Carlo (MCMC) analysis to estimate the model parameters, directly from late time observational data. For this we choose the functional form of the interaction at the outset (through Eq. (\ref{fphi1})).}
\item {As a second example, we give a toy model based on the statefinder parameter which allows us to write Hubble as a function of redshift. This is not as accurate as a direct implementation of MCMC, however, it allows us to solve for the interaction profile as a function of redhshift.}
\end{itemize}

\subsection{Direct Implementation of MCMC and Estimation of model parameters}

To perform the first analysis we consider cosmic chronometer (CC) and the 1048 Supernovae type Ia (SNIa) Pantheon compilation data (SN). The CC data set is composed of $31$ points which are inferred from a differential aging method using passively evolving galaxy pairs separated by small redshift intervals \citep{2014RAA....14.1221Z, Jimenez:2003iv, Moresco:2016mzx, simon, moresco, stern, Moresco:2015cya}. These measurements are independent of Cepheid distance scale measurements and any cosmological model. Despite being somewhat dependent on the modeling of stellar ages, they are based on a robust stellar population synthesis technique \citep{Gomez-Valent:2018hwc, Lopez-Corredoira:2017zfl, Lopez-Corredoira:2018tmn,Verde:2014qea, moresco, Moresco:2016mzx}. The resulting $\chi^2_H$ estimator is given by 
\begin{equation}
    \chi^2_H(\Theta)=\sum_{i=1}^{31}\frac{\left(H(z_i,\,\Theta)-H_\mathrm{obs}(z_i)\right)^2}{\sigma_H^2(z_i)}\,,
\end{equation}
where $H(z_i,\Theta)$ are the theoretical Hubble parameter values at redshift $z_i$ with model parameters $\Theta$, $H_\mathrm{obs}(z_i)$ are the corresponding measured values of the Hubble parameter at $z_i$ with observational error of $\sigma_H(z_i)$. \\

The SN data set is a compilation of 1048 SNIa relative luminosity distance measurements \citep{Pan-STARRS1:2017jku}. This publicly available release of the SN data set is corrected for systematic effects. However, since the apparent magnitude of each SNIa needs to be calibrated through a fiducial absolute magnitude constant $M$, this parameter will appear as a nuisance parameter in the MCMC analyses, which can be implemented through the theoretical values of the distance moduli in a straightforward manner.
\begin{equation}
    \mu(z_i,\,\Theta)=5\log_{10}\left[D_L(z_i,\,\Theta)\right]+M\,,
\end{equation}
for redshifts $z_i$ corresponding to the luminosity distance
\begin{equation}
    D_L(z_i,\Theta)=c\,(1+z_i)\int_0^{z_i}{\frac{\mathrm{d}z'}{H(z',\,\Theta)}}\,,
\end{equation}
where $c$ is the speed of light. The Hubble constant will then be marginalized over the MCMC analyses where the $\chi^2_\mathrm{SN}$ is prescribed by \citep{2011ApJS..192....1C}
\begin{equation}
    \chi_{\text{SN}}^2(\Theta) = \left(\Delta\mu(z_i,\,\Theta)\right)^{T} C_{\text{SN}}^{-1}\, \Delta\mu(z_i,\,\Theta)+\ln\left({\frac{S}{2\pi}}\right)-\frac{k^2(\Theta)}{S}\,,
\end{equation}
where $C_{\text{SN}}^{}$ is the total covariance matrix, $S$ is the sum of all the components of $C_{SN}^{-1}$, while $k$ is given by
\begin{equation}
    k(\Theta)={\left(\Delta\mu(z_i,\,\Theta)\right)^{T}\cdotp C_{\text{SN}}^{-1}}\,,
\end{equation}
with $\Delta\mu(z_i,\,\Theta)=\mu(z_i,\,\Theta)-\mu_{\text{obs}}(z_i)$.

\begin{figure}[h]
\centering
\includegraphics[width=0.48\textwidth]{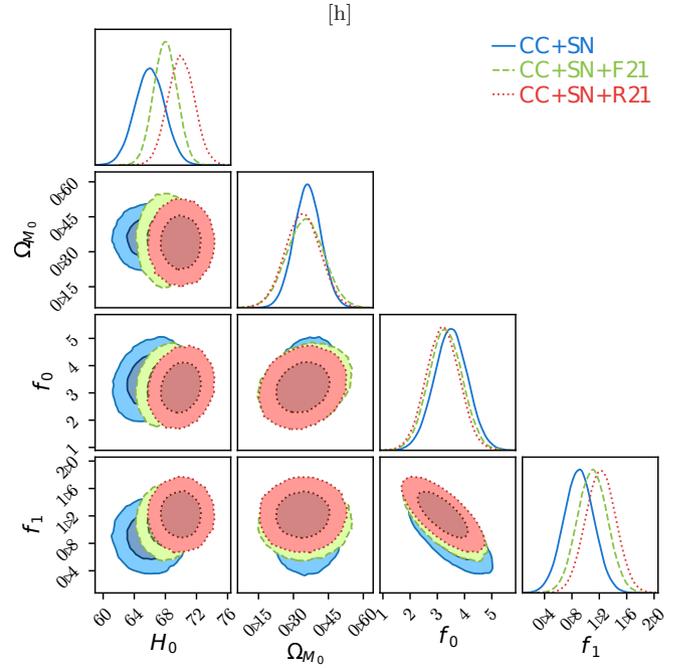}
\caption{Posteriors and confidence levels for the CBD scalar--tensor model using CC+SN data together with the R21 and F21 priors.}
\label{obs_output}
\end{figure}

In addition to the CC and SN data sets, we also consider the impact of priors on the Hubble constant in the CBD MCMC analyses. In the constraint analyses, we consider three scenarios: (i) The combined CC and SN data sets without any prior on $H_0$ which gives the result of purely observational data; (ii) we add the latest SH0ES local estimate \citep{Riess:2021jrx} of $H_0 = 73.04 \pm 1.04 \,{\rm km\, s}^{-1} {\rm Mpc}^{-1}$ (R21) which is based on the Hubble flow together with the combined CC+SN data sets; (iii) we also consider the Hubble constant obtained from the tip of the red giant branch from Ref.~\citep{Freedman:2021ahq} given as $H_0 = 69.8 \pm 1.71 \,{\rm km\, s}^{-1} {\rm Mpc}^{-1}$ (F21) which we again use in conjunction with the combined $CC+SN$ data sets. Other measurements exist in the literature \citep{Abdalla:2022yfr} but the measurements discussed here are the most representative in terms of model-independent local values and are also the most extremes of the reported values of the Hubble constant in the literature.

\begin{table}
    \centering
    \caption{Results for the CBD model MCMC analysis. First column: data sets used together with $H_0$ prior. Second column: $H_0$ values derived from the MCMC analysis. Third column: Constrained values of $\Omega_{m_0}$. Fourth column: Best fit $f_0$ values. Fifth column: Best fit $f_1$ values.}
    \label{tab:results}
    \begin{tabular}{ccccc}
        \hline
		Model & $H_0$ & $\Omega_{M_0}$ & $f_0$ & $f_1$ \\ 
		\hline
		CC+SN & $66.0\pm 1.9$ & $0.360^{+0.057}_{-0.059}$ & $3.50^{+0.62}_{-0.61}$ & $0.92^{+0.20}_{-0.24}$ \\ 
		CC+SN+F21 & $68.0^{+1.5}_{-1.6}$ & $0.352^{+0.079}_{-0.082}$ & $3.27^{+0.65}_{-0.58}$ & $1.11^{+0.21}_{-0.23}$ \\ 
		CC+SN+R21 & $69.8^{+2.0}_{-1.5}$ & $0.340^{+0.075}_{-0.076}$ & $3.20\pm 0.61$ & $1.23^{+0.22}_{-0.23}$ \\ 
		\hline
    \end{tabular}
\end{table}

The results of the MCMC analyses is shown in Fig.~\ref{obs_output} together with the parameter constraints shown in Table~\ref{tab:results} where outputs are shown for $CC+SN$ without a prior, with the F21 prior and then with the R21 prior. The fits give reasonable values of for the Hubble constant and matter density parameter. As expected the impact of priors on the Hubble constant do lead to higher posterior values of the value of $H_0$ while simultaneously producing lower values of $\Omega_{{\rm M}_0}$. As for the model parameters, this hosts a much richer structure than normal with two model parameters for the CBD theory. These parameters appear to be anti-correlated and to be fairly stable under the appearance of priors in the data sets.   \\

\subsection{An Analytical Reconstruction Based on Statefinder}
In this subsection we try to give a simple toy model that can describe late-time cosmology and provide a closed form of the Hubble Function. In the cosmological equation Eq. (\ref{cosmofield}), we have four unknowns, the scale factor $a$, density $\rho_{m}$, the interaction $f(\phi)$ and the scalar field $\phi$, but only three independent equations to solve for them. We essentially need an ansatz over one of the unknowns. Keeping in mind that it is non-trivial to directly find an exact solution of this system of equations, we establish a desirable Hubble form, from a kinematic reconstruction technique. We use the cosmic statefinder parameter, a dimensionless, purely kinematic quantity, written as a combination of the Hubble parameter $H(z)$ and it's derivatives,

\begin{eqnarray}
&& q = -\frac{\ddot{a}a}{\dot{a}^2} = -\frac{\dot H}{H^2}-1, \\&&
r = \frac{\stackrel{\bf{...}}{a}}{aH^3} = \frac{\ddot H}{H^3}+3\frac{\dot H}{H^2} + 1, \label{eq:eq1.3} \\&&
s = \frac{r-1}{3\left(q-\frac{1}{2}\right)}. \label{eq:eq1.4}
\end{eqnarray}

For a detailed analysis of an analytical statefinder reconstruction we refer to \citep{scmnras}. The main essence is using $1+z = \frac{1}{a} \equiv x$ as a variable and writing the definition of statefinder $s$ as a differential equation

\begin{equation}
s(x) = \frac{-2x\frac{H'}{H} + \left(\frac{H'^2}{H^2}+\frac{H''}{H}\right)x^2}{3\left(\frac{H'}{H}x - \frac{3}{2} \right)}.
\end{equation}

If we assume that the statefinder can be rendered as a constant $s = \delta - \frac{2}{3}$ during the deceleration to acceleration transition of the universe at late-times, as a function of redshift the Hubble can be written as

\begin{eqnarray}\nonumber\label{hubblez}
&& H(z) = \\&&\nonumber
\frac{100 h_{0}}{(1+C_{1})^{1/2}}(1+z)^{\frac{1}{4}\left[1+3\delta- \left\lbrace 1 + 6\delta + 9\delta^2 + 36\left(\delta - \frac{2}{3}\right)\right\rbrace^{1/2}\right]} \\&&
\left[C_{1} + (1+z)^{\left\lbrace 1 + 6\delta + 9\delta^2 + 36\left(\delta - \frac{2}{3}\right)\right\rbrace^{1/2}}\right]^{1/2}.
\end{eqnarray}

$C_{1}$ is a constant of integration. $h_0$ is dimensionless, a scaled version of the present value of Hubble parameter, written by dividing $H_0$ with $100$ km $\mbox{Mpc}^{-1}$ $\mbox{sec}^{-1}$. Indeed, this is a resticted scenario, and can be rendered as a special case of perhaps more generalized scenario where $s(x)$ is a function of redshift. From Eq. (\ref{hubblez}), to ensure a real evolution we must also put a constraint on $\delta$ as
\begin{equation}
\delta^2 + \frac{14}{3}\delta - \frac{23}{9} > 0.
\end{equation} 

The analytical form of Hubble enables us to estimate the model parameters from observational data. We take a simple set, a combination of $JLA+OHD+BAO$, to analyze the best fit parameter values and associated uncertainty regions.  \\

\begin{figure}
\begin{center}
\includegraphics[angle=0, width=0.52\textwidth]{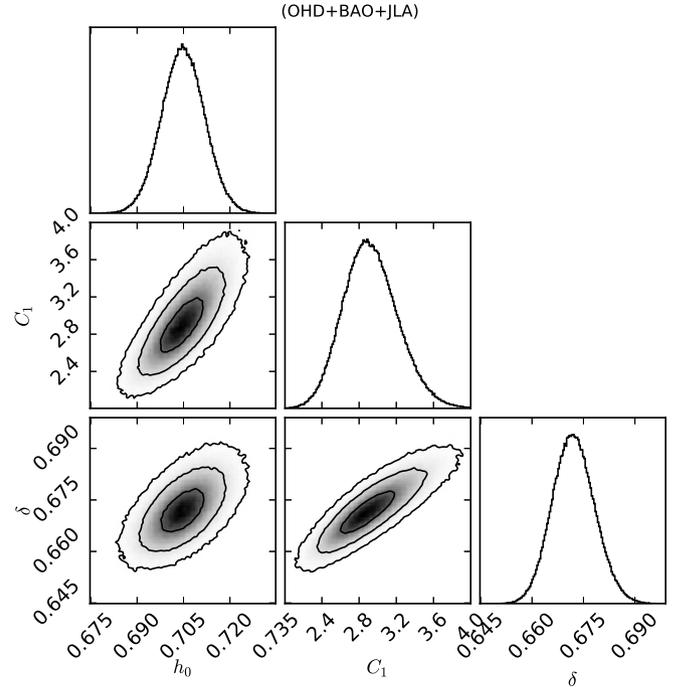}
\caption{Combined analysis of OHD+JLA+BAO : Best fit and associated 1$\sigma$, 2$\sigma$ confidence contours on the parameter space}
\label{Modelcontour}
\end{center}
\end{figure}

The numerical code used in this subsection is a python implementation of MCMC \citep{mcmc}. In Fig. \ref{Modelcontour}, we show the estimated parameters of the model and the likelihood regions through confidence contours. The best-fit values of the model parameters and 1$\sigma$ estimated errors are written in Table \ref{resulttable}. Using the best fit value of $\delta$, the estimated statefinder parameter during late times is $0.011 \geq s \geq -0.001$. \\

\begin{table}
\caption{{\small The parameter values and the associated 1$\sigma$ uncertainty of the parameters, obtained from the analysis with different combinations of the data sets.}}\label{resulttable}
\begin{tabular*}{\columnwidth}{@{\extracolsep{\fill}}lrrrrl@{}}
\hline
 & \multicolumn{1}{c}{$h_0$} & \multicolumn{1}{c}{$C_{1}$} & \multicolumn{1}{c}{$\delta$} \\
\hline
$OHD+JLA+BAO$ 	  & $0.705^{+0.007}_{-0.007}$ &$2.913^{+0.303}_{-0.274}$ & $0.672^{+0.006}_{-0.006}$ &\\
\hline
\end{tabular*}
\end{table}

\begin{figure}
\begin{center}
\includegraphics[angle=0, width=0.40\textwidth]{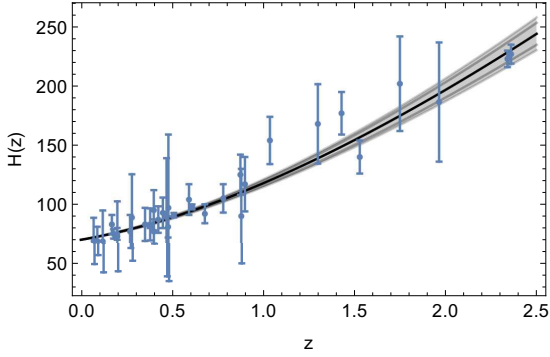}
\caption{Hubble parameter $H(z)$ as a function of redshift alongwith observational data points. The best fit parameter plot is in thick black and associated confidence regions are in gray.}
\label{Hz_data}
\end{center}
\end{figure}

We plot the evolution of $H(z)$ for the reconstructed model in Fig. \ref{Hz_data} and show that for best fit parameter values the Hubble parameter around $z \sim 0$ is very close to astrophysical observations. It shows particular agreement with the $CC+SN+R21$ estimation based on a larger range of observations (See Table. \ref{tab:results}).

\begin{figure}
\begin{center}
\includegraphics[width=0.40\textwidth]{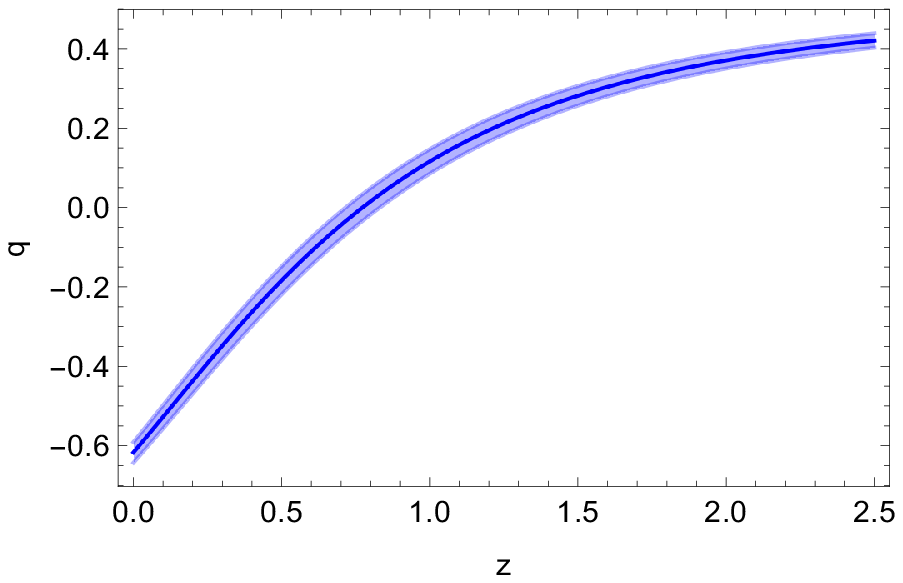}
\includegraphics[width=0.40\textwidth]{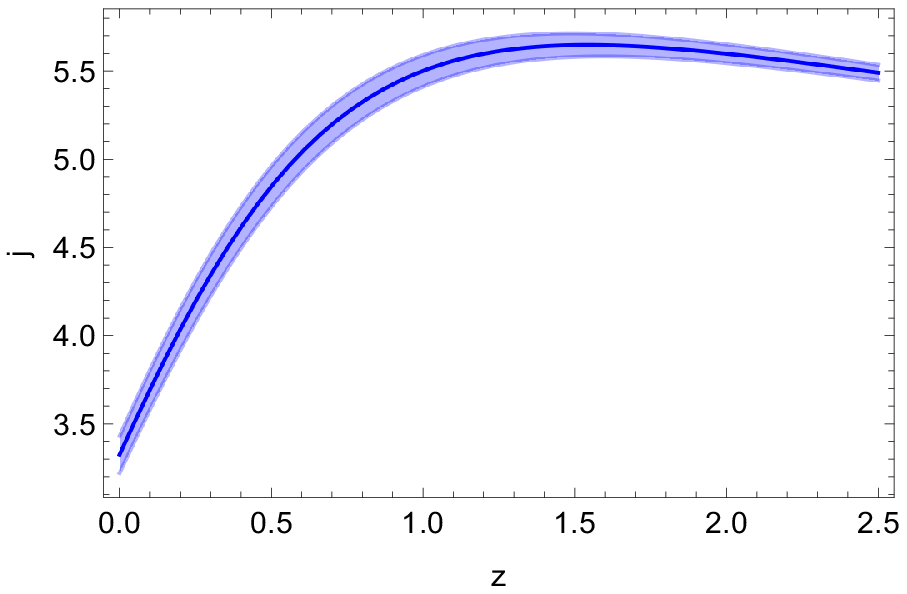}
\caption{Deceleration parameter $q(z)$ (top graph) and the jerk parameter $j(z)$ (bottom graph) vs redshift. The best fit parameter plots are in bold blue and the associated 3$\sigma$ confidence regions are in faded blue.}
\label{kinematic_parameters}
\end{center}
\end{figure}

The deceleration parameter $q(z)$ and the jerk parameter $j(z)$ exhibit interesting non-trivial evolution as a function of redshift as shown in Fig. \ref{kinematic_parameters}. Evolution for the best fit parameter values (bold blue curve) and the allowed departure due to the average uncertainty in parameter estimation (faded blue curve) are shown. The present value of the deceleration parameter is close to $-0.62$, which matches well with observations. The transition in the signature of $q(z)$ marks the transition from decelerated phase into an accelerated phase and the transition redshift $z_{t} < 1$ is also consistent with observations. We also note that the jerk parameter evolves with redshift and the present value is greater than $1$, hinting at a departure from standard $\Lambda$CDM. \\

An additional benefit of having a closed form of Hubble is that we can consider the thermodynamic equilibrium of the cosmological system. We write the total entropy of the universe as a combination of the entropy of cosmological horizon and fluid components, $S = S_f + S_h$ \citep{gibbons, jacobson}. The thermodynamic equilibrium requires that 
\begin{eqnarray}
&& \frac{dS}{dn}\geq 0, \\&&
\frac{d^2S}{dn^2} < 0, \label{therm}
\end{eqnarray}
with $n = \ln{a}$. Using this redefinition of coordinate one can transform the above equations into \citep{jamil}

\begin{eqnarray}\label{Sdn}
&&S_{,n} \propto \frac{(H_{,n})^2}{H^4}, \\&&
S_{,nn} = 2S_{,n}\left(\frac{H_{,nn}}{H_{,n}}-\frac{2H_{,n}}{H}\right) = 2S_{,n}\Psi.
\end{eqnarray}

\begin{figure}
\begin{center}
\includegraphics[width=0.40\textwidth]{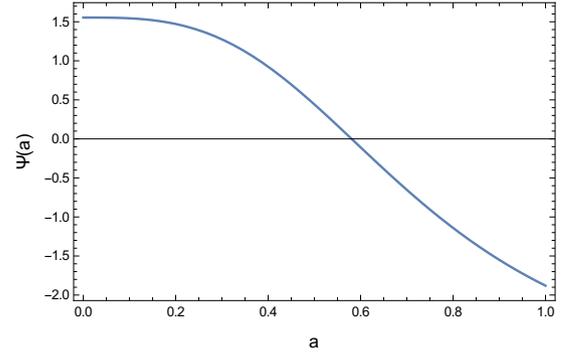}
\caption{$\left(\frac{H_{,nn}}{H_{,n}}-\frac{2H_{,n}}{H}\right)=\Psi$ vs the scale factor $a$. Only best fit parameter values are used, $h_{0} \sim 0.71$, $C_{1} \sim 2.913$ and $\delta \sim 0.67$.}
\label{psiplot}
\end{center}
\end{figure}

Therefore a thermodynamic equilibrium ($S_{,nn} < 0$) requires $\Psi < 0$. We ensure this by drawing $\Psi$ vs $a$ for the kinematic model in Fig. \ref{psiplot}. $\Psi$ remains in a negative domain during the late times and the evolution follows closely a corresponding $\Lambda$CDM behavior, such as a smooth transition from positive into negative values.

\begin{figure}
\begin{center}
\includegraphics[width=0.40\textwidth]{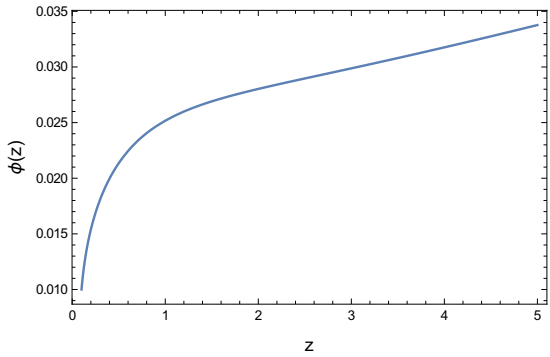}
\caption{Evolution of the Brans Dicke scalar in Jordan Frame as a function of redshift for the best fit parameter values, $h_{0} \sim 0.71$, $C_{1} \sim 2.913$ and $\delta \sim 0.67$.}
\label{BDeinplot}
\end{center}
\end{figure}

Altogether, these establish an observationally viable cosmological behavior and a desired Hubble form. It still requires to reconstruct and solve for the other unknown functions of the theory, such as the BD scalar and it's interaction. We write the components of the field equations as functions of redshift and use the solution for Hubble as in Eq. (\ref{hubblez}) tonumerically solve the field equations. We solve for the best fit parameter values and show the evolution of the chameleon-BD scalar in Fig. \ref{BDeinplot}. The plot suggests that the geometric scalar evolves slowly during deceleration. Since the transition from deceleration into acceleration, the first derivative of the scalar becomes increasingly high and the scalar starts taking a dominating role. \\

\begin{figure}
\begin{center}
\includegraphics[width=0.40\textwidth]{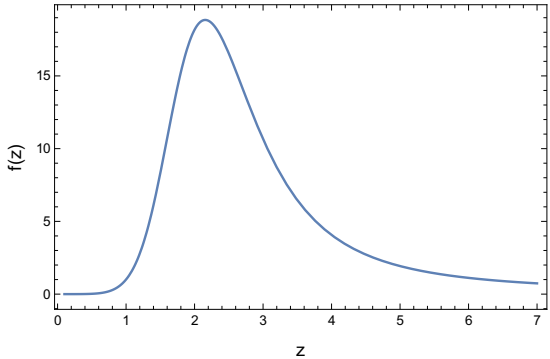}
\caption{Evolution of the Scalar-Matter interaction in Jordan Frame as a function of redshift for the best fit parameter values, $h_{0} \sim 0.71$, $C_{1} \sim 2.913$ and $\delta \sim 0.67$.}
\label{intereinplot}
\end{center}
\end{figure}

In Fig. \ref{intereinplot}, we plot the scalar-matter interaction $f(\phi)$ as a function of $z$. Intriguingly, we note that during a late-time acceleration, the interaction $f(\phi)$ varies very slowly with $z$. Therefore we can speculate that a slowly varying interaction of this scalar field with Baryonic matter, is probably a viable assumption, atleast during late time acceleration. The interaction can also play a crucial role in acting as a switch of smooth transition, since a strong interaction indicates deceleration and a dominating scalar with subdued interaction drives the acceleration.

\begin{figure}
\begin{center}
\includegraphics[width=0.40\textwidth]{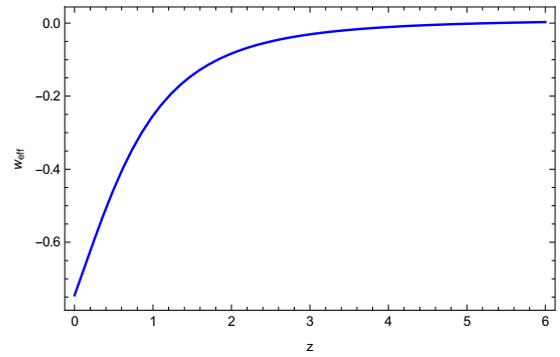}
\caption{Evolution of equation of state in Jordan Frame as a function of redshift for the best fit parameter values for $\omega_{m} = 0$, $h_{0} \sim 0.71$, $C_{1} \sim 2.913$ and $\delta \sim 0.67$.}
\label{EOSeinplot1}
\end{center}
\end{figure}

Due to the numerical solutions we can plot the effective eqation of state (EOS) of the interacting scalar-matter system in Fig. \ref{EOSeinplot1}. The EOS exhibits a dark energy dominated acceleration around $z \sim 0$ ($\omega_{eff} \sim -1$) and a matter dominated deceleration for higher redshifts ($\omega_{eff} \sim 0$).

\begin{figure}
\begin{center}
\includegraphics[width=0.40\textwidth]{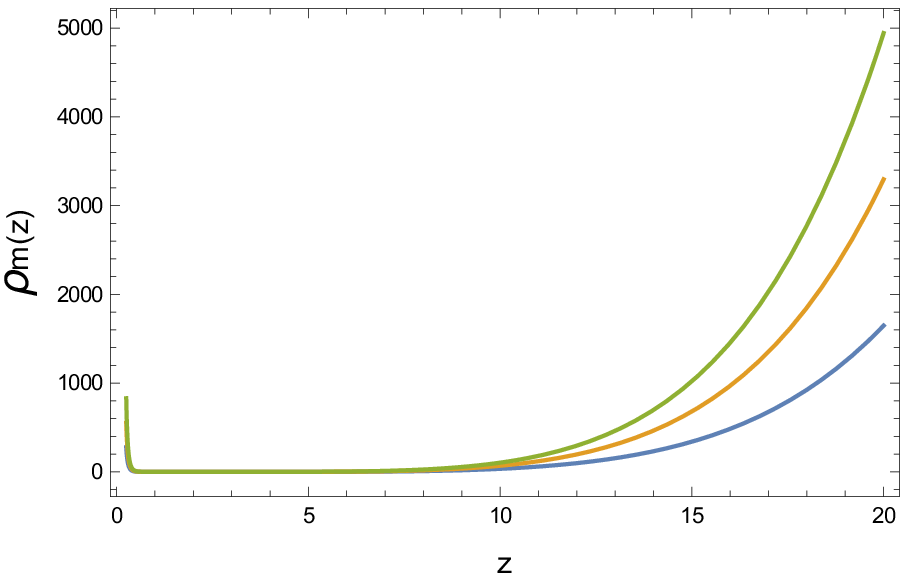}
\includegraphics[width=0.40\textwidth]{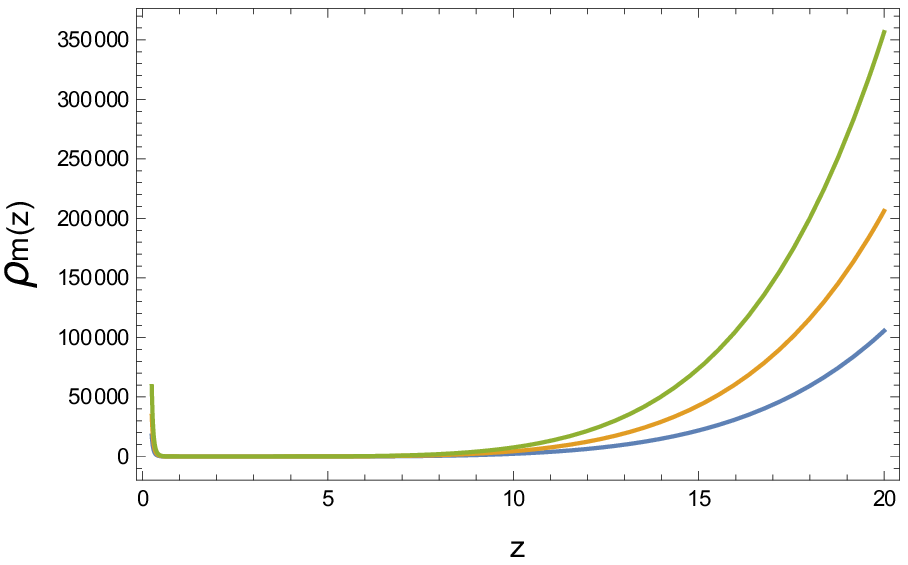}
\caption{Density of the fluid in Jordan Frame as a function of redshift for the best fit parameter values, $h_{0} \sim 0.71$, $C_{1} \sim 2.913$ and $\delta \sim 0.67$. Top Panel : Fluid density for different $\rho_0$. Bottom Panel : Fluid Density for different choices of $\omega_{BD}$.}
\label{EOSeinplot3}
\end{center}
\end{figure}

In Fig. \ref{EOSeinplot3}, we plot the density of the accompanying pressureless fluid as a function of redshift for best fit model parameter values. It depends on the parameter $\rho_{0}$, as well as on the Brans-Dicke parameter. The graph on top is for different values of $\rho_{0}$ while other parameters are kept fixed. The graph on bottom is for different $\omega_{BD}$ for a fixed $\rho_0$. The graphs clearly suggest that the fluid density in the interacting scalar-matter scenario sees a steady decay during the epoch of deceleration, perhaps starting from a preceding era of early acceleration. It becomes completely subdued around the redshift of deceleration-to-acceleration transition. Essentially, this scenario gives us some hints of what an interacting Dark Matter-Dark Energy cosmology might look like, where the scalar field acts as the Dark Energy and the fluid acts as a pressurless Dark matter and the interaction between them plays the role of a trigger of the transition across different epochs. We also comment that the behavior of the fluid density during late-time cosmology is independent on the $BD$ parameter as can be seen from the graph on bottom panel. We also note that there is an increase in the matter energy densities at small redshifts. This may not be entirely physical, however, given the fact that we have a geometric scalar field interacting with Baryonic matter distributions, there may be underlying dynamical phenomenologies involved. The present work is not completely suitable for addressing all such aspects and we intend to extend this further using a dynamical symmetry analysis in a future work. \\ 
 
\begin{figure}
\begin{center}
\includegraphics[width=0.40\textwidth]{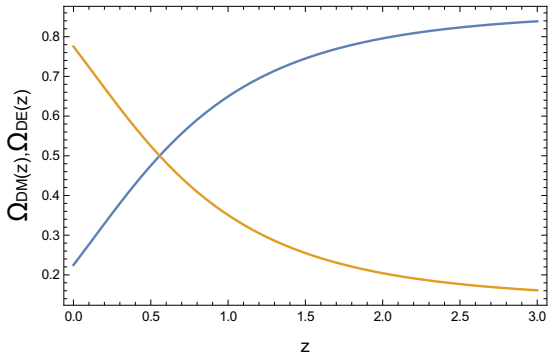}
\caption{Comparison of the evolving Dark Energy density (yellow) with the perfect fluid density (blue) in Jordan Frame as a function of redshift, for the best fit parameter values, $h_{0} \sim 0.71$, $C_{1} \sim 2.913$ and $\delta \sim 0.67$.}
\label{EOSeinplot4}
\end{center}
\end{figure}

Keeping in mind a possible comparison between Dark Energy and Dark Matter distribution, we plot the scaled energy density comparison between these components, $\Omega_{DM}(z)$ and $\Omega_{DE}(z)$ in Fig. \ref{EOSeinplot4}. These components are defined in a manner such that
\begin{eqnarray}\nonumber
&& \Omega_{DM}(z) = \frac{\rho_{DM}}{\rho}, \\&&\nonumber
\Omega_{DM}(z) + \Omega_{DE}(z) = 1.
\end{eqnarray}

The plot clearly suggests that the cosmology under consideration is Dark Energy dominated in present era (yellow curve) and Dark Matter/fluid dominated during the preceding deceleration (blue curve).

\section{Conclusion}
The present work serves as a special generalization of standard Brans-Dicke theory. The geometric scalar field in a standard Brans-Dicke theory modifies the gravitational interaction by making the Newtonian constant a function of coordinates. We further extend the role of this scalar and construct a model where the scalar interacts with ordinary (baryonic) matter and in the process, acquires a density-dependent mass term. This accounts for a chameleon like behavior of the scalar, which is demonstrated for instance by it's evasion from local astronomical tests. \\

We discuss that if the geometric scalar must behave as a chameleon, the extended interaction profiles of the theory must obey some requirements such that the Equivalence Principle (EP) is not violated on the solar system scales. We follow methodically the seminal analysis of Khoury and Weltman \citep{khoury} and deduct a modified set of constraints on the scalar-matter coupling for our extended version. This also leads us to some modified set of constraints over the parameters in the lagrangian of the theory, for instance, the Brans-Dicke parameter. Assuming that the self-interaction potential of the Brans-Dicke scalar is of a runaway form, we ultimately derive that for some special cases, the mass parameter can be comparable with the scale associated with the cosmological constant. We show that with this generalized Chameleon-Brans-Dicke scalar field it is possible to satisfy the astrophysical requirments and describe the smooth transition of the universe from deceleration into acceleration. These requirements of consistency leads to restricted profiles of the scalar and the interaction. This is discussed using a kinematic ansatz over the Hubble evolution and estimation of model parameters with a statistical analysis of observational data sets such as the Joint Light-Curve Analysis (JLA), Hubble parameter estimation (OHD) and the Baryon Accoustic Oscillation (BAO) data. \\

On a local astronomical scale, a standard chameleon scalar must satisfy the mass bound $m_o^{-1} \lesssim \text{Mpc}$, where $m_0$ is the mass of the chameleon field at present cosmological density \citep{wang}. Essentially this can rule out any cosmological potential of a scalar of chameleon nature, atleast in the context of the present acceleration. However, we emphasize that the the present construct of extended scalar-tensor theory, do not have a vanishing divergence of the matter stress-energy tensor at the outset as there is an interaction with standard (baryonic) matter in the Jordan Frame itself. With a modified matter conservation law as well as altered geodesic equations the constraint over the mass of the chameleon field in the present theory is expected to be different. \\

In summary, the generalized Chameleon-Brans-Dicke type theories with generalized scalar interactions serve an overall purpose of promoting the idea that standard $\Lambda$CDM is only the simplest special case of a far more enriched structure of the universe. These theories do rejuvinate interests in generalized scalar-tensor theories where the scalar field(s) can cleverly evade detection from local experiments through some sort of screening mechanisms. In the seminal work of Khoury and Weltmann, the key deduction is that a scalar-matter interaction in Einstein Frame can do the trick and help the scalar in it's successful escapade from observation. The first clear extension we derive is the fact that even with a mild scalar-matter interaction in the Jordan Frame, the geometric scalar remains as good a candidate as it can be as a chameleon. The second result is derived from a cosmological analysis, such that the interacting scalar-matter construction is also a good fit to construct a viable late-time acceleration. Cosmologically this interacting model can also motivate a search for unified Dark Energy-Dark Matter interacting model, where the non-trivial interaction in fact can act as a controller in between deceleration and acceleration. Finally, in view of the fifth force as well as solar system constraints (such as difference in free-fall acceleration within solar system) we derive a mathematcial constraint over the Brans-Dicke parameter which has the potential to resolve it's longstanding conflict with local astronomical tests. The cosmological analysis is primarily based on statistical analysis of data from astrophysical observations and kinematic parameters. The analysis does not require any pre-assigned constraints on the structure of the theory and therefore, in principle, can be equally applied in other extensions of the Brans-Dicke theory for instance \citep{nordtvedt, kofinas}. We expect to report discussions on these extensions in a separate work in near future.

\section*{Acknowledgment}
The authors thank Professor Narayan Banerjee for useful suggestions. SC thanks IISER Kolkata where parts of the research was done. KD is supported in part by the grant $MTR/2019/000395$ and Indo-Russian project grant $DST/INT/RUS/RSF/P-21$, both funded by the DST, Govt of India.

{\bf Data Availability Statement} This manuscript has no associated data or the data will not be deposited.

\appendix\section{Exact Static Solutions for the Chameleon Scalar}
A complete solution to Eq. (\ref{sun2}) requires a matching of two solutions and a continuity of the solution at the origin. There are a few possible cases on which the interior and exterior solutions depend upon.

\subsection{Low-Contrast Solution : $R_c = R$}
A solution in this limit is known as the low-contrast solution \cite{waterhouse}, where Eq. (\ref{outsidesolution}) describes the outside of the sphere and Eq. (\ref{otherharmonicsolution}) describes the inside. For a finite $\varphi\left(r\right)$ at $r\rightarrow 0$, we fix $E = -Fe^{-m_\text{c}R}$ and write Eq. (\ref{otherharmonicsolution}) as
\begin{equation}
\varphi(r) = F\frac{e^{m_\text{c}\left(r-R\right)}-e^{-m_\text{c}\left(r+R\right)}}{r} + \varphi_\text{c}.
\end{equation}
The full solution is written as
\begin{displaymath}
\varphi(r)=\left\{
\begin{aligned}
&F\frac{e^{m_\text{c}\left(r-R\right)}-e^{-m_\text{c}\left(r+R\right)}}{r} + \varphi_\text{c} & r&<R\\
&A\frac{e^{-m_\infty(r-R)}}{r} + \varphi_\infty & r&>R.
\end{aligned}
\right.
\end{displaymath}

The equations for the derivative of $\varphi$ are,
\begin{displaymath}
\frac{d\varphi}{dr}=\left\{
\begin{aligned}
&\frac{F}{r^2}\Big(m_\text{c}re^{m_\text{c}\left(r-R\right)}-e^{m_\text{c}\left(r-R\right)}+m_\text{c}re^{-m_\text{c}\left(r+R\right)}\\
&+e^{-m_\text{c}\left(r+R\right)}\Big), r<R \\
&A\frac{-m_\infty re^{-m_\infty\left(r-R\right)}-e^{-m_\infty\left(r-R\right)}}{r^2}, r>R.
\end{aligned}
\right.
\end{displaymath}

With this, the continuity conditions are written as 
\begin{eqnarray}
&&\lim_{r\rightarrow R^-}\varphi (r) = \lim_{r\rightarrow R^+}\varphi(r), \\&&
\lim_{r\rightarrow R^-}\frac{d\varphi}{dr} = \lim_{r\rightarrow R^+}\frac{d\varphi}{dr}.
\end{eqnarray}
These consition lead to two linear equations for $A$ and $F$. The solutions to these equations are
\begin{eqnarray}\nonumber
&& A = \frac{\varphi_\infty - \varphi_\text{c}}{m_\text{c} + m_\infty + m_\text{c}e^{-2m_\text{c}R} - m_\infty e^{-2m_\text{c}R}}\Big(1-m_\text{c}R \\&&
- e^{-2m_\text{c}R} - m_\text{c}Re^{-2m_\text{c}R}\Big), \\&&\nonumber
F = \frac{\varphi_\infty - \varphi_\text{c}}{m_\text{c} + m_\infty + m_\text{c}e^{-2m_\text{c}R} - m_\infty e^{-2m_\text{c}R}}\Big(1 \\&&
+ m_\infty R\Big).
\end{eqnarray}

\subsection{Thick-Shell Solution : $R_c = 0$}
This limit leads to the so-called thick-shell solution. In this case Eq. (\ref{harmonicsolution}) gives the outside of the sphere and Eq. (\ref{inharmonicsolution}) gives the inside. The continuity condition at the origin requires $C = 0$ in Eq. (\ref{inharmonicsolution}) and this allows us to write
\begin{displaymath}
\varphi(r)=\left\{
\begin{aligned}
\frac{\beta_{0}}{6M_\text{p}}\rho_\text{c}r^2 + D\varphi_\text{c}, \qquad r<R \\
A\frac{e^{-m_\infty(r-R)}}{r} + \varphi_\infty, \qquad r>R,
\end{aligned}
\right.
\end{displaymath}
and
\begin{displaymath}
\frac{d\varphi}{dr}=\left\{
\begin{aligned}
\frac{\beta_{0}}{3M_\text{p}}\rho_\text{c}r, \qquad r<R \\
A\frac{-m_\infty r e^{-m_\infty\left(r-R\right)}-e^{-m_\infty\left(r-R\right)}}{r^2}, \qquad r>R.
\end{aligned}
\right.
\end{displaymath}

The continuity conditions are thereafter solved for the coefficients $A$ and $D$ to write
\begin{eqnarray}
&& A = -\frac{\beta_{0}}{3M_\text{p}}\rho_\text{c}\frac{R^3}{1+m_\infty R},\\&&
D = \frac{\varphi_\infty}{\varphi_\text{c}} - \Big(\frac{1}{1+m_\infty R} + \frac{1}{2}\Big)\frac{\beta_{0}\rho_\text{c}R^2}{3\varphi_\text{c}M_\text{p}}.
\end{eqnarray}

\subsection{Thin-Shell Solution : $0 < R_c < R$}
This is called the thin-shell solution and is of utmost interest to us. The solution is written by an assembly of solutions for three regions, through the respective continuity conditions. Using $E = -Fe^{-m_\text{c}R_\text{c}}$ we write $\varphi$ and $d\varphi/dr$ as
\begin{gather*}
\varphi(r)=\left\{
\begin{aligned}
&F\frac{e^{m_\text{c}\left(r-R_\text{c}\right)}-e^{-m_\text{c}\left(r+R_\text{c}\right)}}{r} + \varphi_\text{c} & r&\in\left(0,R_\text{c}\right)\\
&\frac{\beta_{0}}{6M_\text{p}}\rho_\text{c}r^2 + \frac{C}{r} + D\varphi_\text{c} & r&\in\left(R_\text{c},R\right)\\
&A\frac{e^{-m_\infty(r-R)}}{r} + \varphi_\infty & r&\in\left(R,\infty\right),
\end{aligned}
\right.\\
\frac{d\varphi}{dr}=\left\{
\begin{aligned}
&\frac{F}{r^2}\Big(m_\text{c}r e^{m_\text{c}\left(r - R_\text{c}\right)} - e^{m_\text{c}\left(r-R_\text{c}\right)} + m_\text{c}re^{-m_\text{c}\left(r+R_\text{c}\right)} \\
& +e^{-m_\text{c}\left(r+R_\text{c}\right)}\Big), \qquad\qquad\qquad\qquad r\in\left(0,R_\text{c}\right)\\
&\frac{\beta_{0}}{3M_\text{p}}\rho_\text{c}r - \frac{C}{r^2}, \qquad\qquad\qquad\qquad r\in\left(R_\text{c},R\right)\\
&A\frac{-m_\infty re^{-m_\infty\left(r-R\right)} - e^{-m_\infty\left(r-R\right)}}{r^2}, r\in\left(R,\infty\right).
\end{aligned}
\right.
\end{gather*}

The continuity equations at $R_\text{c}$ are

\begin{eqnarray}\label{firsttwocontinuityequations}
&&F\frac{1-e^{-2m_\text{c}R_\text{c}}}{R_\text{c}}+\varphi_\text{c}=\frac{\beta_{0}}{6M_\text{p}}\rho_\text{c}R_\text{c}^2 + \frac{C}{R_\text{c}} + D\varphi_\text{c},\\&&\nonumber
F\frac{m_\text{c}R_\text{c} - 1 + m_\text{c}R_\text{c}e^{-2m_\text{c}R_\text{c}} + e^{-2m_\text{c}R_\text{c}}}{R_\text{c}^2} = \frac{\beta_{0}}{3M_\text{p}}\rho_\text{c}R_\text{c} \\&&
-\frac{C}{R_\text{c}^2}.
\end{eqnarray}

The continuity equations at $R$ are,
\begin{equation}
\frac{\beta_{0}}{6M_\text{p}}\rho_\text{c}R^2 + \frac{C}{R} + D\varphi_\text{c} = A\frac{1}{R}+\varphi_\infty,\\
\label{thirdcontinuityequation}
\end{equation}
\begin{equation}
\frac{\beta_{0}}{3M_\text{p}}\rho_\text{c}R-\frac{C}{R^2} = A\frac{-m_\infty R-1}{R^2}.
\label{fourthcontinuityequation}
\end{equation}

The solution of the continuity equations depend on $R_\text{c}\in\left[0,R\right]$, which is to be chosen in a manner such that the harmonic approximation in Eq. (\ref{harmonicapproximation}) is a better approximation for $r\in\left(0,R_\text{c}\right)$ and the approximation in Eq. (\ref{linearapproximation}) is better for $r\in\left(R_\text{c},R\right)$. The thick-shell solution is applicable for $R_\text{c}=0$, otherwise, $R_\text{c}$ is defined by
\begin{displaymath}
m_\text{c}^2 \left(\varphi\left(R_\text{c}\right) - \varphi_\text{c}\right) = \frac{\beta_{0}}{M_\text{p}}\rho_\text{c},
\end{displaymath}
and the scalar $\varphi\left(r\right)$ is given by the thin-shell case. The $r\rightarrow R_\text{c}^-$ limit of the thin-shell solution gives
\begin{gather}
m_\text{c}^2 F\frac{1-e^{-2m_\text{c}R_\text{c}}}{R_\text{c}} = \frac{\beta_{0}}{M_\text{p}}\rho_\text{c}\notag\\
\Rightarrow F = \frac{\beta_{0}\rho_\text{c}R_\text{c}}{m_\text{c}^{2} M_\text{p}\left(1-e^{-2m_\text{c}R_\text{c}}\right)}.
\label{R0equation}
\end{gather}

Essentially one determines $F$ in terms of $R_\text{c}$ and other parameters from the continuity equations using Eq. (\ref{R0equation}). This is quite comparable to standard Chameleon screening, for instance, in the limit $m_\infty R\ll1$ and $F = 0$, $\left(R-R_\text{c}\right)/R\ll1$, in thin shell case. The low-contrast solution is the $R_\text{c}\rightarrow R$ limit of the thin-shell solution and is not of that interest to us. The exterior approximate solution for all the cases is 
\begin{displaymath}
\varphi\left(r\right) = A\frac{e^{-m_\infty(r-R)}}{r} + \varphi_\infty.
\end{displaymath}
We note that the dimensionless constant $A$ gives us the strength of the chameleon force due to the scalar and we shall discuss the structure of this constant in brief. For a thick-shell approximation, assuming $m_\infty R\ll1$, we can write
\begin{align}
A&=-\frac{\beta_{0}}{3M_\text{p}}\rho_\text{c}\frac{R^3}{1+m_\infty R}\\
&\approx-\frac{\beta_{0}}{4\pi M_\text{p}}\left(\frac{4}{3}\pi R^3\rho_\text{c}\right).
\end{align}

However, for a thin-shell approximation and $F = 0$, the two continuity equations in Eq. (\ref{firsttwocontinuityequations}) give
\begin{gather}
C = \frac{\beta_{0}}{3M_\text{p}}\rho_\text{c}R_\text{c}^3,\\
D = 1-\frac{\beta_{0}\rho_\text{c}R_\text{c}^2}{2M_\text{p}}\frac{1}{\varphi_\text{c}}.
\end{gather}
On substituting into the continuity Eq. (\ref{fourthcontinuityequation}), we get
\begin{gather}
\frac{\beta_{0}}{3M_\text{p}}\rho_\text{c}R - \frac{\beta_{0}}{3M_\text{p}}\rho_\text{c}\frac{R_\text{c}^3}{R^2}\approx-\frac{A}{R^2}\notag\\
\Rightarrow A\approx-\frac{\beta_{0}}{3M_\text{pl}}\rho_\text{c}\left(R^3-R_\text{c}^3\right).
\label{Aapproximation}
\end{gather}
Again, substituting into Eq. (\ref{thirdcontinuityequation}) gives
\begin{gather*}
\frac{\beta_{0}}{2M_\text{p}}\rho_\text{c}R^2 + \varphi_\text{c} - \frac{\beta_{0}\rho_\text{c}R_\text{c}^2}{2M_\text{p}} = \varphi_\infty\\
\Rightarrow R^2 - R_\text{c}^2 = \frac{2M_\text{p}}{\beta_{0}\rho_\text{c}}\left(\varphi_\infty - \varphi_\text{c}\right).
\end{gather*}
Finally, we can Taylor-expand Eq. (\ref{Aapproximation}) in $R_\text{c}$ about $R$ to get
\begin{align}\label{thinshellfactor}
A&\approx-\frac{\beta_{0}}{3M_\text{p}}\rho_\text{c}\frac{3}{2}R\left(R^2 - R_\text{c}^2\right)\\
&=-\frac{\beta_{0}}{3M_\text{p}}\rho_\text{c}\frac{3}{2}R\frac{2M_\text{p}}{\beta_{0}\rho_\text{c}}\left(\varphi_\infty - \varphi_\text{c}\right)\\
&=-\frac{\beta_{0}}{4\pi M_\text{p}}\left(\frac{4}{3}\pi R^3\rho_\text{c}\right)\frac{3M_\text{p}\left(\varphi_\infty - \varphi_\text{c}\right)}{\beta_{0}\rho_\text{c}R^2}.
\end{align}
The approximate external solutions are therefore written as
\begin{eqnarray}
&&\varphi_\text{thick}\left(r\right) \sim - \frac{\beta_{0}}{4\pi M_\text{p}}\left(\frac{4}{3}\pi R^3 \rho_\text{c}\right)\frac{e^{-m_\infty(r-R)}}{r} + \varphi_\infty \\&&\nonumber
\varphi_\text{thin}\left(r\right) \sim - \frac{\beta_{0}}{4\pi M_\text{pl}}\left(\frac{4}{3}\pi R^3\rho_\text{c}\right)\left(3\frac{M_\text{p}\left(\varphi_\infty - \varphi_\text{c}\right)}{\beta_{0}\rho_\text{c}R^2}\right) \\&&
\frac{e^{-m_\infty(r-R)}}{r} + \varphi_\infty.
\end{eqnarray}


\begin{thebibliography}{99}

\bibitem[\protect\citeauthoryear{Abdalla et. al.}{2022}]{Abdalla:2022yfr}
Abdalla, E. et. al., 2022, FERMILAB-CONF-22-192-SCD, 3, arXiv : astro-ph.CO 2203.06142.

\bibitem[\protect\citeauthoryear{Adak and Dutta}{2014}]{dutta2}
Adak D. and Dutta K., 2014, Phys. Rev. D. 90, no.4, 043502.

\bibitem[\protect\citeauthoryear{Ade et al.}{2014}]{ade}
Ade P. A. R. et al., 2014, Astron. Astrophys. 571, A16.

\bibitem[\protect\citeauthoryear{Adelberger, Heckel and Nelson}{2003}]{adelberger}
Adelberger E. G., Heckel B. R. and Nelson A. E., 2003, Ann. Rev. Nucl. Part. Sci. 53, 77.

\bibitem[\protect\citeauthoryear{Anderson et al.}{2012}]{anderson}
Anderson L. et al., 2012, Mon. Not. R. Astron. Soc. 441, 24.

\bibitem[\protect\citeauthoryear{Arkani-Hamed, Motl, Nicolis and Vafa}{2007}]{arkani}
Arkani-Hamed N., Motl L., Nicolis A. and Vafa C., 2007, JHEP 06, 060.

\bibitem[\protect\citeauthoryear{Baessler et al.}{1999}]{baessler}
Baessler S., Heckel B. R., Adelberger E. G., Gundlach J. H., Schmidt U. and Swanson H. E., 1999, Phys. Rev. Lett. 83, 3585.

\bibitem[\protect\citeauthoryear{Banerjee and Sen}{1997}]{nb1}
Banerjee N. and Sen S., 1997, Phys. Rev. D 56, 1334.

\bibitem[\protect\citeauthoryear{Banerjee and Pavon}{2001}]{nb2}
Banerjee N. and Pavon D., 2001, Phys. Rev. D 63, 043504.

\bibitem[\protect\citeauthoryear{Banerjee and Pavon}{2001}]{nb3}
Banerjee N. and Pavon D., 2001, Class. Quant. Gravit. 18, 593.

\bibitem[\protect\citeauthoryear{Banks, Dine, Fox and Gorbatov}{2003}]{banks}
Banks T., Dine M., Fox P. J. and Gorbatov E., 2003, JCAP 06, 001. 

\bibitem[\protect\citeauthoryear{Barker}{1978}]{barker}
Barker B. M., 1978, Astrophys. J. 219, 5.

\bibitem[\protect\citeauthoryear{Barreiro, Carlos and Copeland}{1998}]{barreiro}
Barreiro T., de Carlos B. and Copeland E. J., 1998, Phys. Rev. D. 57, 7354.

\bibitem[\protect\citeauthoryear{Barrow}{1990}]{barrow}
Barrow J. D., 1990, Phys. Lett. B. 235, 40.

\bibitem[\protect\citeauthoryear{Bergmann}{1968}]{bergmann}
Bergmann P. G., 1968, Int. J. Theor. Phys. 1, 25.

\bibitem[\protect\citeauthoryear{Bernardo and Said}{2021}]{Bernardo:2021qhu}
Bernardo, R. C. and Levi Said, J., 2021, JCAP, 09, 014.

\bibitem[\protect\citeauthoryear{Bertolami and Martins}{2000}]{bertolami}
Bertolami O. and Martins P. J., 2000, Phys. Rev. D 61, 064007.

\bibitem[\protect\citeauthoryear{Bertolami, Boehmer, Harko and Lobo}{2007}]{bertolami1}
Bertolami O., Bohmer C. G., Harko T. and Lobo F. S. N., Phys. Rev. D. 75, 104016.

\bibitem[\protect\citeauthoryear{Bertolami and Paramos}{2010}]{bertolami2}
Bertolami O. and Paramos J., 2010, J. Phys. Conf. Ser. 222, 012010.

\bibitem[\protect\citeauthoryear{Bertolami and Martins}{2012}]{bertolami3}
Bertolami O. and Martins A., 2012, Phys. Rev. D. 85, 024012.

\bibitem[\protect\citeauthoryear{Bertolami, Frazao and Paramos}{2011}]{bertolami4}
Bertolami O., Frazao P. and Paramos J., 2011, Phys. Rev. D. 83, 044010.

\bibitem[\protect\citeauthoryear{Betoule et al.}{2014}]{betoule}
Betoule M. et al., 2014, Astron. Astrophys. 568, A22. 

\bibitem[\protect\citeauthoryear{Beutler et al.}{2011}]{beutler}
Beutler F. et. al., 2011, Mon. Not. R. Astron. Soc. 416, 3017. 

\bibitem[\protect\citeauthoryear{Binetruy, Gaillard and Wu}{1997}]{binetruy}
Binetruy P., Gaillard M. K. and Wu Y. Y., 1997, 412(3-4), 288.

\bibitem[\protect\citeauthoryear{Blake et al.}{2012}]{blake}
Blake C. et. al., 2012, Mon. Not. R. Astron. Soc. 425, 405.

\bibitem[\protect\citeauthoryear{Brans and Dicke}{1961}]{bransdicke}
Brans C. and Dicke R. H., 1961, Phys. Rev., 124, 925.

\bibitem[\protect\citeauthoryear{Brax et. al.}{2004}]{brax}
Brax P. et al., 2004, Phys. Rev. D 70, 123518.

\bibitem[\protect\citeauthoryear{Brown}{1993}]{brown}
Brown J. D., 1993, Class. Quant. Gravit. 10, 1579.

\bibitem[\protect\citeauthoryear{Chakrabarti}{2021}]{scmnras}
Chakrabarti S., 2021, Mon. Not. Roy. Astron. Soc., 502 (2), 1895 ; Mon. Not. Roy. Astron. Soc., 506 (2), 2518.

\bibitem[\protect\citeauthoryear{Chuang and Wang}{2013}]{chuang}
Chuang C. H. and Wang Y., 2013, Mon. Not. R. Astron. Soc. 435, 255.

\bibitem[\protect\citeauthoryear{Conley et. al.}{2011}]{2011ApJS..192....1C}
Conley, A. et. al., 2011, ApJS, 192(1), 1.

\bibitem[\protect\citeauthoryear{Copeland, Sami and Tsujikawa}{2006}]{copeland}
Copeland E., Sami M. and Tsujikawa S., 2006, Int. J. Mod. Phys. D, 15, 1753.

\bibitem[\protect\citeauthoryear{Damour and Polyakov}{1994}]{damour}
Damour T. and Polyakov A. M., 1994, Nucl. Phys. B 423, 532 ; Gen. Rel. Grav. 26, 1171.

\bibitem[\protect\citeauthoryear{Das, Corasaniti and Khoury}{2006}]{das}
Das S., Corasaniti P. S. and Khoury J., Phys. Rev. D. 73, 083509.

\bibitem[\protect\citeauthoryear{Das and Banerjee}{2008}]{dasbanerjee}
Das S. and Banerjee N., 2008, Phys. Rev. D 78, 043512.

\bibitem[\protect\citeauthoryear{Delubac et al.}{2015}]{delubac}
Delubac T. et al., 2015, Astron. Astrophys. 574, A59.

\bibitem[\protect\citeauthoryear{Dutta and Sorbo}{2007}]{dutta1}
Dutta K. and Sorbo L., 2007, Phys. Rev. D 75, 063514.

\bibitem[\protect\citeauthoryear{Easson}{2007}]{easson}
Easson D. A., 2007, JCAP, 0702, 004.

\bibitem[\protect\citeauthoryear{Ellis, Kalara, Olive and Wetterich}{1989}]{ellis}
Ellis J., Kalara S., Olive K. A. and Wetterich C., 1989, Phys. Lett. B 228, 264.

\bibitem[\protect\citeauthoryear{Faraoni}{1999}]{faraoni1}
Faraoni V., 1999, Phys. Rev. D 59, 084021.

\bibitem[\protect\citeauthoryear{Faraoni}{1999}]{faraoni2}
Faraoni V. and Gunzig E., 1999, Int. J. Theor. Phys. 38, 217.

\bibitem[\protect\citeauthoryear{Faraoni}{2004}]{faraoni}
Faraoni V., 2004, Cosmology in Scalar-Tensor Gravity (Kluwer Academic Publishers, Dordrecht, The Nether-
lands).

\bibitem[\protect\citeauthoryear{Faraoni}{2009}]{faraonimatter}
Faraoni V., 2009, Phys. Rev. D. 80, 124040.

\bibitem[\protect\citeauthoryear{Freedman}{2021}]{Freedman:2021ahq}
Freedman, W. L., 2021, Astrophys. J., 919(1), 16.

\bibitem[\protect\citeauthoryear{Frieman, Hill, Stebbins and Waga}{1995}]{frieman}
Frieman J. A., Hill C. T., Stebbins A. and Waga I., 1995, Phys. Rev. Lett. 75, 2077.

\bibitem[\protect\citeauthoryear{Foreman-Mackey et al.}{2013}]{mcmc}
Foreman-Mackey D. et al. 2013, Astron. Soc. Pac. 125, 306.

\bibitem[\protect\citeauthoryear{Gibbons and Hawking}{1977}]{gibbons}
Gibbons G. W. and Hawking S. W., 1977, Phys. Rev. D, 15, 2738.

\bibitem[\protect\citeauthoryear{G\'omez-Valent and Amendola}{2018}]{Gomez-Valent:2018hwc}
G\'omez-Valent, A. and Amendola, L., 2018, JCAP, 04, 051.

\bibitem[\protect\citeauthoryear{Gubser and Khoury}{2004}]{gubser}
Gubser S. S. and Khoury J., 2004, Phys. Rev. D 70, 104001.

\bibitem[\protect\citeauthoryear{Guth}{1981}]{guth}
Guth A., 1981, Phys. Rev. D 23, 347.

\bibitem[\protect\citeauthoryear{Halverson et. al.}{2002}]{halverson}
Halverson N. W. et al., 2002, Astrophys. J., 568, 38.

\bibitem[\protect\citeauthoryear{Hill and Ross}{1988}]{hill}
Hill C. T. and Ross G. C., 1988, Nucl. Phys. B 311, 253.

\bibitem[\protect\citeauthoryear{Hinterbichler and Khoury}{2010}]{hinterbichler}
Hinterbichler K. and Khoury J., 2010, Phys. Rev. Lett. 104, 231301.

\bibitem[\protect\citeauthoryear{Huey, Steinhardt, Ovrut and Waldrum}{2000}]{huey}
Huey G., Steinhardt P. J., Ovrut B. A. and Waldram D., 2000, Phys. Lett. B 476, 379.

\bibitem[\protect\citeauthoryear{Jacobson}{1995}]{jacobson}
Jacobson T., 1995, Phys. Rev. Lett., 75, 1260.

\bibitem[\protect\citeauthoryear{Jaffe et. al.}{2001}]{jaffe}
Jaffe A. H. et al., 2001, Phys. Rev. Lett., 86, 3475.

\bibitem[\protect\citeauthoryear{Jain and Khoury}{2010}]{jain}
Jain B. and Khoury J., 1010, Annals Phys. 325, 1479.

\bibitem[\protect\citeauthoryear{Jamil, Saridakis and Setare}{2010}]{jamil}
Jamil M., Saridakis E. N. and Setare M. R., 2010, JCAP, 1011, 032.

\bibitem[\protect\citeauthoryear{Jarv, Kuusk, Saal and Vilson}{2015}]{jarv}
Jarv L., Kuusk P., Saal M. and Vilson O., 2015, Phys. Rev. D 91, 024041.

\bibitem[\protect\citeauthoryear{Jimenez, Verde, Treu and Stern}{2003}]{Jimenez:2003iv}
Jimenez, R., Verde, L., Treu, T. and Stern, D., 2003, Astrophys. J., 593, 622.

\bibitem[\protect\citeauthoryear{Khoury and Weltman}{2004}]{khoury}
Khoury J. and Weltman A., 2004, Phys. Rev. Lett. 93, 171104 ; Phys. Rev. D 69, 044026.

\bibitem[\protect\citeauthoryear{Kofinas, Papantonopoulo and Saridakis}{2016}]{kofinas}
Kofinas G., Papantonopoulos E. and Saridakis E. N., 2016 Class. Quant. Grav. 33, 155004.

\bibitem[\protect\citeauthoryear{La and Steinhardt}{1989}]{la}
La D. and Steinhardt P. J., 1989, Phys. Rev. Lett. 62, 376.

\bibitem[\protect\citeauthoryear{Lange et. al.}{2001}]{lange}
Lange A. E. et al., 2001, Phys. Rev. D, 63, 042001.

\bibitem[\protect\citeauthoryear{Lopez-Corredoira, Vazdekis, Gutierrez, and Castro-Rodriguez}{2017}]{Lopez-Corredoira:2017zfl}
Lopez-Corredoira, M., Vazdekis, A., Gutierrez, C. M. and Castro-Rodriguez, N., 2017, Astron. Astrophys., 600, A91.

\bibitem[\protect\citeauthoryear{Lopez-Corredoira and Vazdekis}{2018}]{Lopez-Corredoira:2018tmn}
Lopez-Corredoira, M. and Vazdekis, A., 2018, Astron. Astrophys., 614, A127.

\bibitem[\protect\citeauthoryear{Mathiazhagan and Johri}{1984}]{mathia}
Mathiazhagan C. and Johri V. B., 1984, Class. Quant. Grav. 1, L29.

\bibitem[\protect\citeauthoryear{Melchiorri et. al.}{2000}]{melchi}
Melchiorri A. et al., 2000, Astrophys. J. Lett., 536, L63.

\bibitem[\protect\citeauthoryear{Misner, Thorne and Wheeler}{1973}]{misner}
Misner C., Thorne K. and Wheeler J., 1973, Gravitation (W. H. Freeman and Company, San Francisco.

\bibitem[\protect\citeauthoryear{Moresco et al.}{2012}]{moresco}
Moresco M. et. al., 2012, JCAP 07, 053.

\bibitem[\protect\citeauthoryear{Moresco et al.}{2016}]{Moresco:2016mzx}
Moresco, M. et. al., 2016, JCAP, 05, 014.

\bibitem[\protect\citeauthoryear{Moresco}{2015}]{Moresco:2015cya}
Moresco, M., 2015, Mon. Not. Roy. Astron. Soc. 450(1), L16.

\bibitem[\protect\citeauthoryear{Mota and Shaw}{2006}]{mota1}
Mota D. F. and Shaw D. J., 2006, Phys. Rev. Lett. 97, 151102. 

\bibitem[\protect\citeauthoryear{Mota and Shaw}{2007}]{mota2}
Mota D. F. and Shaw D. J., 2007, Phys. Rev. D 75, 063501.

\bibitem[\protect\citeauthoryear{Netterfield et. al.}{2002}]{netterfield}
Netterfield C. B. et al., 2002, Astrophys. J., 571, 604.

\bibitem[\protect\citeauthoryear{Noller and Nicola}{2019}]{Noller:2018wyv}
Noller, J. and Nicola, A., 2019, Phys. Rev. D. 99(10), 103502.

\bibitem[\protect\citeauthoryear{Nordtvedt Jr.}{1970}]{nordtvedt}
Nordtvedt Jr. K., 1970, Astrophys. J. 161, 1059.

\bibitem[\protect\citeauthoryear{Padmanabhan and Roychoudhury}{2003}]{paddy1}
Padmanabhan T. and Roychoudhury T., 2003, Mon. Not. R. Astron. Soc., 344, 823.

\bibitem[\protect\citeauthoryear{Roychoudhury and Padmanabhan}{2005}]{paddy2}
Roychoudhury T. and Padmanabhan T., 2005, Astron. Astrophys., 429, 807.

\bibitem[\protect\citeauthoryear{Perlmutter et. al.}{1997}]{perl1}
Perlmutter S. et al, 1997, Bull. Am. Astron. Soc., 29, 1351 ; 1999, Astrophys. J., 517, 565.

\bibitem[\protect\citeauthoryear{Quiros et al.}{2015}]{quiros}
Quiros I., Garcia-Salcedo R., Gonzalez T. and Antonio Horta-Rangel F., 2015, Phys. Rev. D. 92, 044055.

\bibitem[\protect\citeauthoryear{Riess et. al.}{1998}]{riess1}
Riess A. G. et al, 1998, Astron. J., 116, 1009. 

\bibitem[\protect\citeauthoryear{Riess}{2001}]{riess2}
Riess A. G., 2001, Astrophys. J., 560, 49.

\bibitem[\protect\citeauthoryear{Riess et. al.}{2021}]{Riess:2021jrx}
Riess, A. G. et al., 2021, arXiv : astro-ph.CO 2112.04510.

\bibitem[\protect\citeauthoryear{Sahni and Starobinsky}{2000}]{sahni}
Sahni V. and Starobinski A. A., 2000, Int. J. Mod. Phys. D, 9, 373.

\bibitem[\protect\citeauthoryear{Sahni}{2004}]{sahninotes}
Sahni V., 2004, Lect. Notes Phys. 653 : 141.

\bibitem[\protect\citeauthoryear{Schutz}{1970}]{schutz}
Schutz B., 1970, Phys. Rev. D. 2, 2762.

\bibitem[\protect\citeauthoryear{Scolnic et. al.}{2018}]{Pan-STARRS1:2017jku}
Scolnic, D. M. et. al., 2018, Astrophys. J., 859(2), 101.

\bibitem[\protect\citeauthoryear{Sergijenko, Durrer and Novosyadlyj}{2011}]{2011JCAP...08..004S}
Sergijenko O., Durrer R. and Novosyadlyj B., 2011, JCAP, 2011(8), 004.

\bibitem[\protect\citeauthoryear{Sen and Sen}{2001}]{sensen}
Sen S. and Sen A. A., 2001, Phys. Rev. D 63, 124006.

\bibitem[\protect\citeauthoryear{Simon, Verde and Jimenez}{2005}]{simon}
Simon J., Verde L. and Jimenez R., 2005, Phys. Rev. D. 71, 123001. 

\bibitem[\protect\citeauthoryear{Sotiriou}{2014}]{soti}
Sotiriou T., 2014, Gravity and Scalar Fields, Based on a Lecture Given at the Seventh Aegean Summer
School Beyond Einsteins Theory of Gravity, arXiv:1404.2955v1 [gr-qc].

\bibitem[\protect\citeauthoryear{Sotiriou and faraoni}{2008}]{sotifara}
Sotiriou T. P. and Faraoni V., 2008, Class. Quant. Gravit. 25, 205002.

\bibitem[\protect\citeauthoryear{Stern et al.}{2010}]{stern}
Stern D. et. al., 2010, JCAP 02, 008.

\bibitem[\protect\citeauthoryear{Tonry et. al.}{2003}]{tonry}
Tonry J. L. et al, 2003, Astrophys. J., 594, 1.

\bibitem[\protect\citeauthoryear{Upadhye, Gubser and Khoury}{2006}]{upadhye}
Upadhye A., Gubser S. S. and Khoury J., 2006, Phys. Rev. D 74, 104024.

\bibitem[\protect\citeauthoryear{Verde, Protopapas and Jimenez}{2014}]{Verde:2014qea}
Verde, L., Protopapas, P. and Jimenez, R., 2014, Phys. Dark Univ., 5, 307.

\bibitem[\protect\citeauthoryear{Wagoner}{1970}]{wagoner}
Wagoner R. V., 1970, Phys. Rev. D 1, 3209. 

\bibitem[\protect\citeauthoryear{Wang, Hui and Khoury}{2012}]{wang}
Wang J., Hui L. and Khoury J., 2012, Phys. Rev. Lett. 109, 241301.

\bibitem[\protect\citeauthoryear{Waterhouse}{2006}]{waterhouse}
Waterhouse T. P., 2006, arXiv:astro-ph/0611816v1.

\bibitem[\protect\citeauthoryear{Will}{2001}]{will1}
Will C. M., 2001, Liv. Rev. Rel. 4, 4.

\bibitem[\protect\citeauthoryear{Will}{2005}]{will2}
Will C. M., 2005, Liv. Rev. Rel. 9, 3.

\bibitem[\protect\citeauthoryear{Yang et. al.}{2019}]{Yang:2018xah}
Yang, W., Shahalam, M., Pal, B., Pan, S. and Wang, A., 2019, Phys. Rev. D., 100(2), 023522.

\bibitem[\protect\citeauthoryear{Zhang et. al.}{2014}]{2014RAA....14.1221Z}
Zhang, C., Zhang, H., Yuan, S., Liu, S., Zhang, T. and Sun, Y., 2014, Research in Astronomy and Astrophysics 14(10), 1221.

\bibitem[\protect\citeauthoryear{Zimdahl and Pavon}{2004}]{zimdahl}
Zimdahl W. and Pavon D., 2004, Gen. Rel. Grav. 36, 1483.

\bibitem[\protect\citeauthoryear{Zlatev, Wang and Steinhardt}{1999}]{zlatev}
Zlatev I., Wang L. and Steinhardt P. J., 1999, Phys. Rev. Lett. 82, 896 ; Phys. Rev. D. 59, 12350.








\end{thebibliography}
\end{document}